\documentclass[usenatbib]{mn2e}
\usepackage{graphicx}
\usepackage{fixltx2e}

\begin{document}
\topmargin-1cm

\newcommand\approxgt{\mbox{$^{>}\hspace{-0.24cm}_{\sim}$}}
\newcommand\approxlt{\mbox{$^{<}\hspace{-0.24cm}_{\sim}$}}
\newcommand{\lexp}{\mathop{\langle}}
\newcommand{\rexp}{\mathop{\rangle}}
\newcommand{\rexpc}{\mathop{\rangle_c}}
\newcommand{\pcl}{pseudo-$C_{l}$~}
\newcommand{\pclc}{Pseudo-$C_{l}$~}
\newcommand{\ylm}{Y_{lm}}
\newcommand{\plm}{\psi_{lm}}
\newcommand{\thi}{\hat{\theta}_{i}}
\newcommand{\thj}{\hat{\theta}_{j}}
\newcommand{\nbar}{\bar{n}}
\newcommand{\talm}{\tilde{a}_{lm}}
\newcommand{\tcl}{\tilde{C}_{l}}
\newcommand{\nhat}{\hat{\bf n}}
\newcommand{\kvec}{{\bf k}}
\newcommand{\khat}{\hat{\bf k}}
\def\bi#1{\hbox{\boldmath{$#1$}}}

\title{The real-space clustering of luminous red galaxies around $z<0.6$
quasars in the Sloan Digital Sky Survey}
\author[Padmanabhan et al.]{Nikhil Padmanabhan$^{1,5}$, Martin White$^{2}$, 
Peder Norberg$^{3}$, Cristiano Porciani$^{4}$ \\
$^{1}$ Physics Division, Lawrence Berkeley National Laboratory,
1 Cyclotron Rd., Berkeley, CA 94720, USA. \\
$^{2}$ Department of Physics and Astronomy, 601 Campbell Hall,
University of California Berkeley, CA 94720, USA.\\
$^3$The Scottish Universities Physics Alliance, 
Institute for Astronomy, University of Edinburgh, \\ Royal Observatory, Blackford Hill, Edinburgh, EH9 3HJ, UK. \\
$^{4}$ Institute for Astronomy, ETH Zurich, 8093 Zurich, Switzerland \\
$^{5}$ Hubble Fellow, Chamberlain Fellow \\
}

\date{\today}
\maketitle

\begin{abstract}
We measure the clustering of a sample of photometrically selected luminous red
galaxies around a low redshift ($0.2<z<0.6$) sample of quasars selected from
the Sloan Digital Sky Survey Data Release 5.
We make use of a new statistical estimator to obtain precise measurements of
the LRG auto-correlations and constrain halo occupation distributions for
them.  These are used to generate mock catalogs which aid in interpreting
our quasar-LRG cross correlation measurements.
The cross correlation is well described by a power law with slope $1.8\pm0.1$
and $r_0=6\pm0.5\,h^{-1}$Mpc, consistent with observed galaxy correlation
functions.  We find no evidence for `excess' clustering on $0.1\,$Mpc scales
and demonstrate that this is consistent with the results of \citet{Ser06} and
\citet{Str07}, when one accounts for several subtleties in the interpretation
of their measurements.
Combining the quasar-LRG cross correlation with the LRG auto-correlations,
we determine a large-scale quasar bias $b_{\rm QSO} = 1.09\pm0.15$ at a median
redshift of $0.43$, with no observed redshift or luminosity evolution.
This corresponds to a mean halo mass
$\langle M\rangle\sim 10^{12}\,h^{-1} M_{\odot}$,
Eddington ratios from 0.01 to 1 and lifetimes less than $10^{7}\,$yr.
Using simple models of halo occupation, these correspond to a number density
of quasar hosts greater than $10^{-3} h^{3} {\rm Mpc}^{-3}$ and stellar masses 
less than $10^{11} h^{-1} M_\odot$.
The small-scale clustering signal can be interpreted with the aid of our mock
LRG catalogs, and depends on the manner in which quasars inhabit halos.
We find that our small scale measurements are inconsistent with quasar
positions being randomly subsampled from halo centers above a mass threshold,
requiring a satellite fraction $>$ 25 per cent.
\end{abstract}


\section{Introduction}
\label{sec:introduction}

Quasars are among the most luminous astrophysical objects, and are believed
to be powered by accretion onto supermassive black holes
\citep[e.g.][]{Sal64,Lyn69}.
They have become a key element in our current paradigm of galaxy
evolution -- essentially all
spheroidal systems at present harbor massive black holes \citep{KorRic95},
the masses of which are correlated with many properties of their host systems.
However, the physical mechanisms that trigger and fuel quasars are still 
unknown; furthermore, it is possible that very different mechanisms 
dominate at low and high redshifts and at high and low luminosities.

Deep imaging with the Hubble Space Telescope 
\citep[e.g.][]{Bah97, McL99, Dun03, Flo04, Let08}
suggest that low-$z$ QSOs are exclusively hosted
by bright galaxies with $L>L_{\star}$.  Radio-loud QSOs reside in
early-type galaxies, while their radio-quiet counterparts have both early and
late type hosts, with the fraction of early type hosts increasing with the quasar
optical luminosity.
In most cases there is strong observational evidence for the presence of a
young (subdominant) stellar population \citep{San04} and large
amounts of gas, irrespective of the morphological type of the host
\citep{Let07}. Furthermore,  30-50 per cent of quasars appear
to be associated with interactions, although the number of such imaged systems 
is small, and signatures of mergers are notoriously difficult to observe.
The emerging picture is that QSO activity and star formation are inextricably
linked \citep[e.g.][]{Sil07,Nan07} in galaxies that contain a massive bulge (and thus a massive
black hole) and a gas reservoir.

The clustering of quasars as a function of redshift and luminosity provides a
different perspective on the above picture.  
The amplitude of clustering on large
scales is related to the masses of the dark matter halos which host the
quasars (their environment), which together with the observed number density
allows us to constrain the quasar lifetimes or duty cycles.
The small-scale clustering of quasars can shed light on the triggering
mechanism for quasars, and the nature of quasar progenitors.

However it is only recently that samples of quasars have grown big enough
(in terms of the number of objects) to study their clustering with some
precision \citep{PorMagNor04,Cro05,PorNor06,Hen06,Mye07a,Mye07b,She07,Ang08}.
One of the major problems with measuring the clustering of quasars is that
they are extremely rare ($\bar{n}\sim 10^{-6}\,h^3\,{\rm Mpc}^{-3}$ at
$z\sim 0.5$).
Shot-noise from Poisson fluctuations in the counts of objects thus
obscures their clustering signal.
At low redshifts, this problem is exacerbated, requiring
measurements in very broad redshift intervals.

To avoid this,  we cross-correlate 
approximately $2,500$ low $z$ quasars \citep{Sch07} with a sample
of $450,000$ luminous red galaxies (LRGs) \citep{Pad07}, both selected from the 
Sloan Digital Sky Survey \citep[SDSS]{York00}
and with overlapping redshift distributions.
The LRGs have very reliable photometric redshifts \citep{Pad05},
trace the matter distribution in a way that is well understood and have
a much higher volume density ($\bar{n}\sim 10^{-4}\,h^3\,{\rm Mpc}^{-3}$)
than the quasar sample.
The cross-correlation can thus be well measured and inverted, using the known
redshift distribution, to the underlying 3D clustering.

In this paper, we make use of several novel techniques for measuring the
clustering of galaxies and quasars, and compute full covariance matrices
for our estimators from the data themselves.  
While the idea of enhancing the clustering signal by using cross-correlations
is not new \citep{Cro04,AdeSte05a,AdeSte05b,Ser06,DEEP2,Str07,Mou08},
the sample size and ability
to perform such detailed statistical analyses is new to this paper.
In addition, the precise measurements of LRG clustering allow us for the first time 
to discuss the manner in which both LRGs and quasars inhabit 
dark matter halos at $z\sim 0.5$.

The outline of the paper is as follows.  In \S\ref{sec:data} we describe
the LRG and quasar samples, drawn from the SDSS, that
we use.  The clustering measurements are described in \S\ref{sec:clustering},
where we pay special attention to the techniques used and the error estimates.
The implications of our results for
quasars is explored in \S\ref{sec:interp}, including comparisons with earlier
work.  In particular we investigate the manner in
which quasars inhabit dark matter halos at $z\sim 0.5$.
We conclude in \S\ref{sec:conclusions}. Appendix \ref{appendix:halomodel}
contains the technical details of the halo model fits used in this paper, 
while Appendix~\ref{appendix:serber} 
recasts the measurements of \citet{Ser06} and \citet{Str07} into the framework
of this paper, highlighting unappreciated subtleties in their interpretation.
Where necessary we shall assume a $\Lambda$CDM cosmological model with
$\Omega_{\rm mat}=0.25$, $\Omega_\Lambda=0.75$, $\sigma_8=0.8$. Also, unless the
$h$ dependence is explicitly specified, we assume $h=0.7$.

\section{Data}
\label{sec:data}

\subsection{Quasars}

\begin{figure}
\begin{center}
\resizebox{3in}{!}{\includegraphics{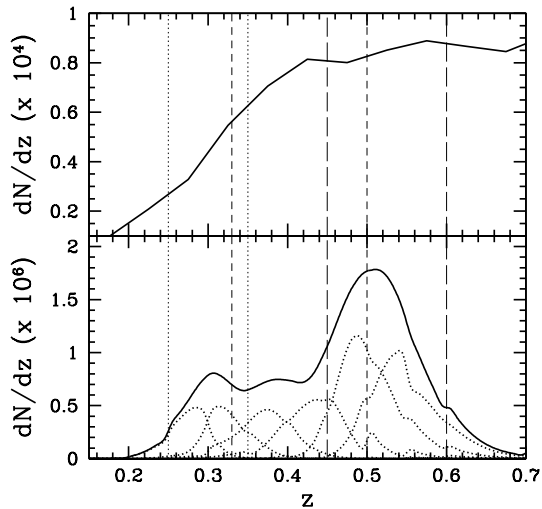}}
\end{center}
\caption{The redshift distribution of the quasars [top] and LRGs [bottom] 
used in this analysis. The LRG redshift distributions are derived from the 
observed photometric redshift distribution, after deconvolving the redshift
errors. The dotted lines show the redshift distribution for the six 
individual $dz_{\rm photo}=0.05$ LRG samples. The vertical lines mark the 
boundaries of the three quasar redshift slices we consider -- $0.25 < z < 0.35$ (dotted),
$0.33 < z < 50$ (short-dashed) and $0.45 < z < 0.6$ (long-dashed).}
\label{fig:dndz}
\end{figure}

\begin{figure}
\begin{center}
\leavevmode
\includegraphics[width=3.0in]{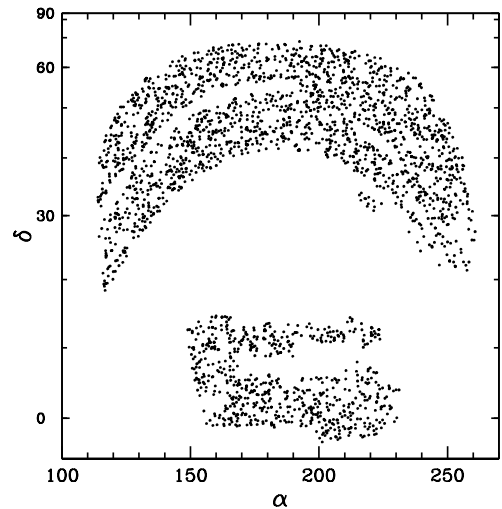}
\end{center}
\caption{The angular distribution of our quasar sample, plotted in an 
RA-$\cos(\delta)$ equal-area rectilinear projection. The angular mask 
for the quasars is both determined by the spectroscopic coverage, as well
as the overlap with the photometric LRG sample.
}
\label{fig:areal_qso}
\end{figure}

We use quasars selected from the fourth edition of the Sloan Digital
Sky Survey (SDSS) quasar catalog \citep{Sch07}. This catalog consists of 
spectroscopically identified quasars in the fifth SDSS data release 
\citep{DR5}, with an absolute PSF magnitude in the $i$ band,
$M_{i} < -22.0$ and at least one emission line with FWHM larger than 
$1000\,{\rm km}{\rm s}^{-1}$.  It does not contain Type 2 QSOs, Seyferts
or BL Lac objects.
In order to construct a homogeneous sample, we follow \citet{Ric06} and
select objects
\begin{itemize}
    \item that were targeted for science ($\rm{SCIENCEPRIMARY = 1}$).
    \item classified by the SDSS photometric pipeline as primary 
    ($\rm{PRIMARY = 1}$).
    \item morphologically consistent with being point sources 
    ($\rm{MORPHOLOGY = 0}$).
    \item with dust and emission line K-corrected $i$-band magnitudes,
    $i<19.1$.
\end{itemize}
In order to cross-correlate with the LRG sample
described below, we restrict ourselves to quasars that lie within the LRG angular 
mask, and with redshifts between $0.25 < z < 0.6$.
The resulting sample (denoted ALL below) has 2476 quasars. 
One subtlety with the SDSS quasar samples are the changes to the 
quasar target selection algorithm \citep{Ric02} over the lifetime of the survey.
However, these changes were made to optimize target selection at high redshifts, and
have little effect on our sample. As we discuss below, 
restricting to quasars selected with the final version of 
the QSO target selection (v3\_1\_0) does not affect any of our 
results. We therefore do not make a cut based on the target selection 
algorithm. 
The redshift distribution is shown in Fig.~\ref{fig:dndz}, 
while the angular distribution is in Fig.~\ref{fig:areal_qso}.

\begin{figure}
\begin{center}
\resizebox{3in}{!}{\includegraphics{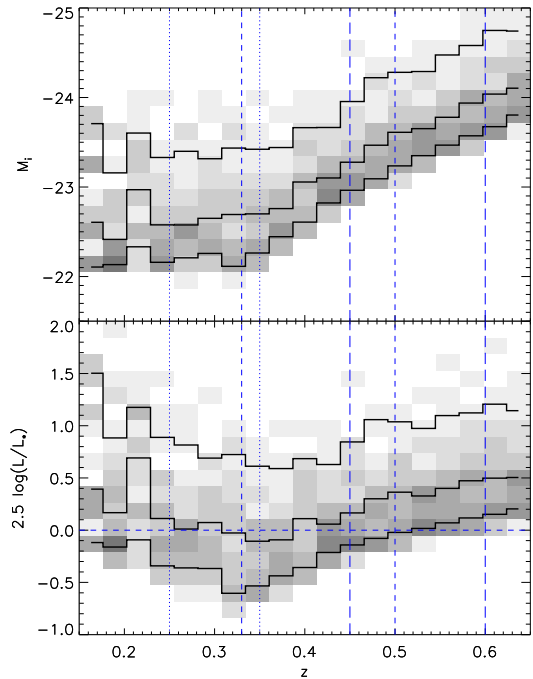}}
\end{center}
\caption{The conditional distribution of absolute magnitude with redshift for
our sample of QSOs. The lines plot the 16, 50 and 84 per cent contours.
The lower panel plots the absolute magnitude relative to $M_\star$, estimated
{}from the 2dF and 2SLAQ survey.
The flattening/upturn at low redshift is due to the $M_i=-22.0$ cut on the
sample, to minimize contamination from the host galaxy.
The vertical lines (as in Fig.~\ref{fig:dndz}) show the redshift boundaries of our samples.}
\label{fig:zL}
\end{figure}

Fig.~\ref{fig:zL} plots the conditional magnitude distribution of our
sample. We assume a redshift evolution of $M_{*,i}$ (k-corrected to $z=0$)
given by
\begin{equation}
  M_{*,i}(z) = M_{*,i,0} - 2.5 (k_{1} z + k_{2}z^{2}) \,\,,
\label{eq:mstar}
\end{equation}
with $M_{*,i,0}=-21.678$, $k_1=1.39$ and $k_2=-0.29$ \citep{2SLAQ},
where we have converted from the SDSS $g$ to $i$ band with $g-i=0.068$
\citep{Ric06}.
This defines a sample of quasars (denoted LSTAR) with $M_i<M_{*,i}$,
yielding an approximately volume limited sample over the redshift 
range we consider.
We can estimate the number density of these samples by integrating the
broken power-law fit to the QSO luminosity function from \citet{2SLAQ},
\begin{equation}
\label{eq:powerlawLF}
\Phi(L,z) = \frac{\Phi^{*}}{10^{0.4(\alpha+1)(M_{i}-M_{*,i})} 
    + 10^{0.4(\beta+1)(M_{i}-M_{*,i})}} \,\,.
\end{equation}
If we assume the parameters estimated from the 2QZ and 2SLAQ surveys 
i.e. $M_{*,i}$ as defined above, $\alpha=-3.31$, $\beta=-1.45$, and
$\Phi^{*}=5.33 \times 10^{-6}$ $h^{3} {\rm Mpc}^{-3} {\rm mag}^{-1}$ \citep{2SLAQ},
and no scatter between $g$ and $i$ magnitudes, we estimate a number 
density of $1.7 \times 10^{-7}$ $h^{3} {\rm Mpc}^{-3}$. If one adopts 
the parameters from \citet{Boy00}, we find a number density of 
$\sim1.3 \times 10^{-7}$ $h^{3} {\rm Mpc}^{-3}$, approximately
$20$ per cent lower.
Note that we keep the sample definition the same for both these cases, 
so one is integrating over magnitudes less than $M_{*}$ defined by
Eq.~\ref{eq:mstar}.

We estimate the bolometric luminosity using the relation from \citet{Cro05},
\begin{equation}
\label{eq:lboldef}
  M_{i} = -2.66 \log_{10}(L_{\rm bol}) + 79.36 \,\,,
\end{equation}
where $L_{\rm bol}$ is in Watts, and we convert from the $b_J$ to $i$
band using the empirical relations in \citet{Ric06}.
For the LSTAR sample at the median redshift of $0.43$, with
$M_{*,i}(z=0.43) = -23.04$, this implies a bolometric luminosity 
$L_{\rm bol} > 10^{38.5}\,$W.
Assuming that the quasars are radiating at the Eddington rate 
($L_{\rm Edd} = 10^{39.1} (M_{\rm bh}/10^{8} M_{\odot})\,$W),
this implies black hole masses $M_{\rm bh} > 3 \times 10^7 M_\odot$. 

Table~\ref{tab:summary} summarizes the various subsamples we consider in this paper
(discussed further in Sec.~\ref{sec:qsocluster}).

\begin{table}
\begin{center}    
\begin{tabular}{lcr}
\hline
Sample & Redshift & $n_{\rm qso}$ \\
\hline
ALL0 & $0.25 < z < 0.35$ & 435 \\
ALL1 & $0.33 < z < 0.50$ & 1277 \\
ALL2 & $0.45 < z < 0.60$ & 1269 \\
\hline
LSTAR0 & $0.25 < z < 0.35$ & 212 \\
LSTAR1 & $0.33 < z < 0.50$ & 751 \\
LSTAR2 & $0.45 < z < 0.60$ & 1094 \\
\hline
\end{tabular}
\caption{\label{tab:summary} A summary of the quasar we
consider.  The columns list the sample name, redshift range, and the number
of quasars.}
\end{center}
\end{table}

\subsection{Luminous red galaxies}

We cross correlate the above quasars with a sample of luminous red galaxies
(LRGs) selected from the SDSS imaging data. The sample selection, angular
mask, and redshift distributions have been described in detail in \cite{Pad07},
and we refer the reader to the details there.
These galaxies have well characterized photometric redshifts and errors
($\delta z \sim 0.03$), allowing us to deconvolve the photometric redshift
distribution to obtain the underlying $dN/dz$ \citep{Pad05}.
We consider LRGs with $0.25 < z_{\rm photo} < 0.55$,
trimmed to an angular mask that covers 3528.04 deg.$^{2}$ in the northern
Galactic hemisphere; this results in a sample of 454882 LRGs.
We then divide the sample into six redshift ranges, summarized in
Table \ref{tab:lrg}, of photometric redshift width $\Delta z=0.05$, where the redshift boundaries
are chosen to select approximately homogeneous samples as the 4000 \AA\ break
shifts from the $g$ to the $r$-band. We found that the LRG properties varied
significantly over the redshift range making fine $z_{\rm photo}$ bins essential
for proper modeling, the choice of $\Delta z=0.05$ being determined by the
photometric redshift errors.
The LRG samples are summarized in Table~\ref{tab:lrg}, 
while the deconvolved redshift distributions are in
plotted in Fig.~\ref{fig:dndz}.

\begin{table*}
\begin{center}
\begin{tabular}{cccccccccc} \hline
Sample & $n_{\rm LRG}$ & $z_{\rm photo}$-range & $\langle z\rangle$ & $z_{\rm mode}$ & $\delta z$ &
 $D$ & $\bar{n}$ & $b_{\rm const}$ & $b_{\rm halo}$ \\ \hline
LRG1 & 29,660 & $0.25<z<0.30$ & 0.276 & 0.287 & 0.029 & 0.87 & 4.7
  & $1.75\pm0.05$ & $1.71\pm0.05$ \\
LRG2 & 32,527 & $0.30<z<0.35$ & 0.326 & 0.312 & 0.033 & 0.86 & 4.1
  & $1.77\pm0.06$ & $1.77\pm0.05$ \\
LRG3 & 41,051 & $0.35<z<0.40$ & 0.376 & 0.375 & 0.049 & 0.84 & 3.4
  & $2.36\pm0.05$ & $2.15\pm0.07$ \\ 
LRG4 & 60,294 & $0.40<z<0.45$ & 0.445 & 0.452 & 0.058 & 0.81 & 3.7
  & $2.28\pm0.05$ & $2.09\pm0.05$ \\
LRG5 & 104,131 & $0.45<z<0.50$ & 0.506 & 0.488 & 0.048 & 0.79 & 4.7
  & $2.02\pm0.04$ & $1.90\pm0.04$ \\
LRG6 & 95,605 & $0.50<z<0.55$ & 0.552 & 0.541 & 0.051 & 0.78 & 4.2
  & $1.90\pm0.05$ & $1.76\pm0.05$ \\
\hline
\end{tabular}
\end{center}
\caption{Properties of the LRG samples, showing the number, photo-$z$ range, mean
and modal redshift, width, growth factor (normalized to unity today) at the
modal redshift, number density (in $10^{-4}\,h^3{\rm Mpc}^{-3}$) and
large-scale bias.  We estimate the large-scale bias by fitting to
$\omega(\theta_s)$ assuming scale-independent bias or fitting to a halo model
(see text).  The quoted errors are purely statistical.  The biases estimated
{}from the different methods have systematic errors at the 5 per cent level.
We adopt $b_{\rm halo}$ as our fiducial value.}
\label{tab:lrg}
\end{table*}

\section{Clustering}
\label{sec:clustering}

\subsection{LRG Clustering}
\label{sec:lrgcluster}

\subsubsection{Methods}

We measure the clustering of the LRG sample in each of the six photometric
redshift slices using the angular clustering estimator described in 
\citet{PWE07}. We define 
\begin{equation}
\label{eq:omegath1}
  \omega(\theta_{s}) \equiv 2\pi \int_{0}^{\theta_{s}}
  \theta\,d\theta\ G(\theta,\theta_{s}) w(\theta) \,\,,
\end{equation}
where
\begin{eqnarray}
\theta_s^3 G(\theta, \theta_{s}) = &
  (x^{2})^{2} \left(1 - x^{2}\right)^{2} \left(\frac{1}{2} - x^{2}\right)\,\, & x<1, \nonumber \\
  = & 0 \,\, & x \ge 1 \,,
\label{eq:omega_Gdef}
\end{eqnarray}
with $x = \theta/\theta_{s}$.
As was shown in \citet{PWE07}, this estimator partially deprojects the 
angular correlation function, yielding a robust estimate of the 
3D real space correlation function on scales of $\sim \frac{1}{2} \bar{\chi} \theta_{s}$
where $\bar{\chi}$ is the mean comoving distance to the redshift slice 
under consideration.

We implement the above estimator using $DD/RR - 1$ as our estimate of the
angular correlation function. Although the \citet{LanSza93} estimate is a
more traditional choice, the contiguous wide-area coverage of the SDSS
imaging makes it unnecessary in our case, and we choose the simpler estimator.
Substituting this into Eq.~\ref{eq:omegath1}, we obtain,
\begin{equation}
\label{eq:omegath2}
  \omega(\theta_{s}) = 2\pi \int \theta\,d\theta
  \ G(\theta,\theta_{s}) \frac{DD}{RR} \,\,,
\end{equation}
where we use the fact that the area-weighted integral of $G(x)$ vanishes by 
construction.  In order to proceed, we note that the $RR$ term is a purely
geometric term determined by the survey mask.  Although this is traditionally
estimated by measuring random-random pairs in the same binning as the $DD$
pairs, \citet{PWE07} point out that on scales much smaller than the size of
the survey, $RR$ is described by 
\begin{equation}
\label{eq:rr}
  RR \propto 2 \pi \theta \Delta\theta\, \Phi(\theta) \,\,,
\end{equation}
where $\Phi(\theta)$ is a smooth function; we obtain a good fit to $\Phi$
using a fifth-order polynomial. 
Having fit $\Phi$, we can make our $\theta$ bins arbitrarily small without
incurring any Poisson noise penalty.
This allows us to rewrite Eq.~\ref{eq:omegath2} as a weighted sum over pairs,
\begin{equation}
\label{eq:omegath3}
\omega(\theta_{s}) = \sum_{i \in DD}
 \frac{G(\theta_{i},\theta_{s})}{\Phi(\theta_{i})} 
 \Theta(\theta_{s}-\theta) \,\,,
\end{equation}
where $\Theta$ is the Heaviside step function.

We estimate the covariance matrix of our measurements by bootstrap resampling
\citep[e.g.][]{Efron}.
An important advantage of $\omega(\theta_{s})$ is its insensitivity to
clustering on scales $\approxgt 2\theta_{s}$.  This allows us to subdivide
the survey into 41 approximately filled spherical rectangles, 7.5$^{\circ}$
in the $RA$ direction and $0.15$ in the $\sin(\delta)$ direction.
Computing $\omega(\theta_{s})$ for each of these subsamples yields 41
independent, identically distributed realizations of $\omega(\theta_{s})$.
The independence of the subsamples is a direct consequence of the 
estimator; this is not true for the more traditional $w(\theta)$.
We then estimate both the average and covariance matrix by bootstrap
resampling these 41 realizations.
In order to improve the numerical stability of this procedure, we
scale $\omega$ by $\theta_s^2$, thereby removing the artificially large
condition number of the covariance matrix that arises due to the large
dynamic range of $\omega$.
The resulting covariance matrix is very well-behaved, with no anomalously
small modes that need be removed.  Note that this is not the case for
$w(\theta)$ which is sensitive to large scale fluctuations, which in turn
lead to unphysical modes in the covariance matrix that must be further
conditioned.  The insensitivity of $\omega(\theta_s)$ to long wavelengths
is an important advantage when attempting to estimate the covariance
matrix from the data itself.

\subsubsection{Results}

\begin{figure*}
\begin{center}
\leavevmode
\includegraphics[width=2.0in]{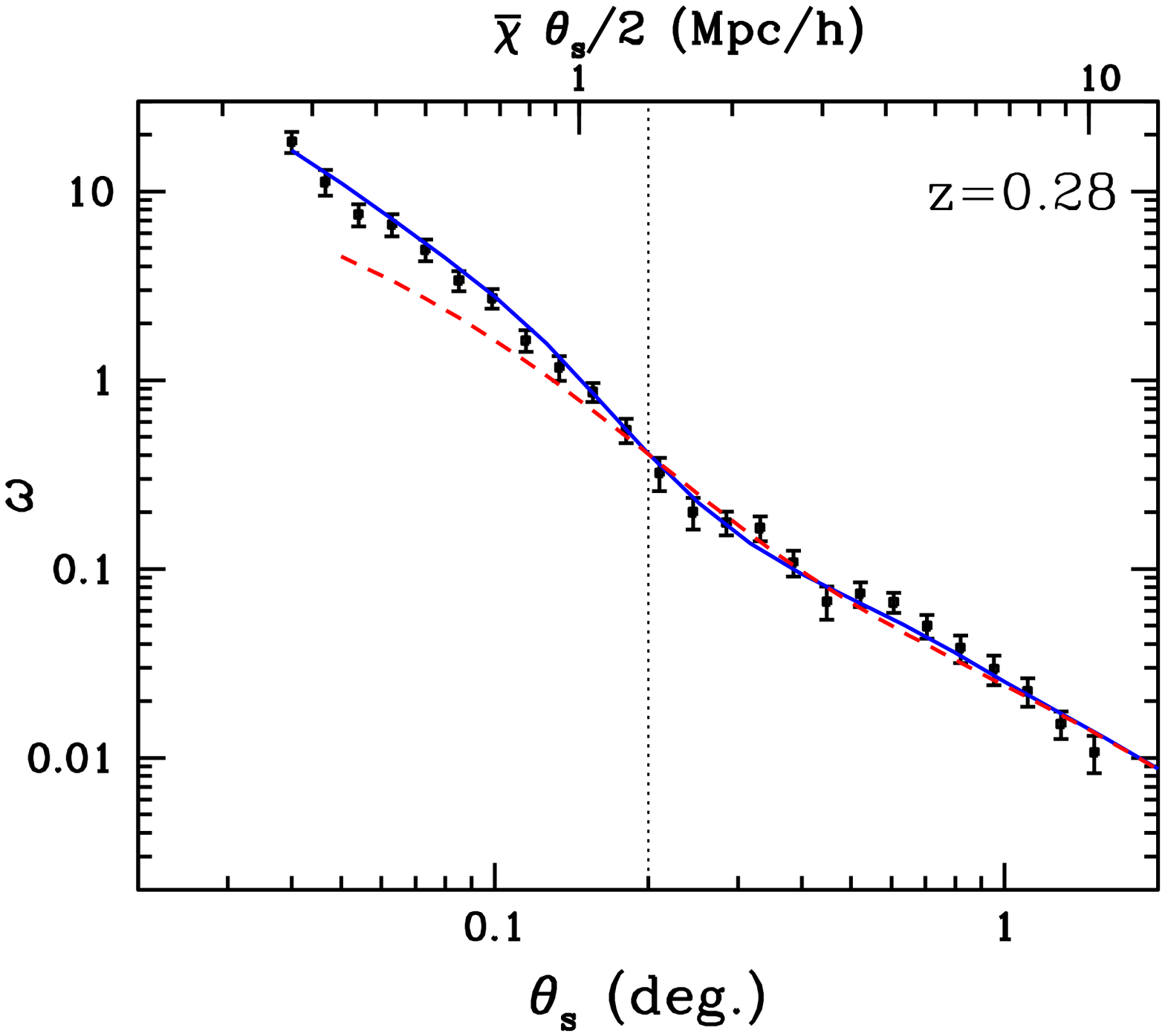}
\includegraphics[width=2.0in]{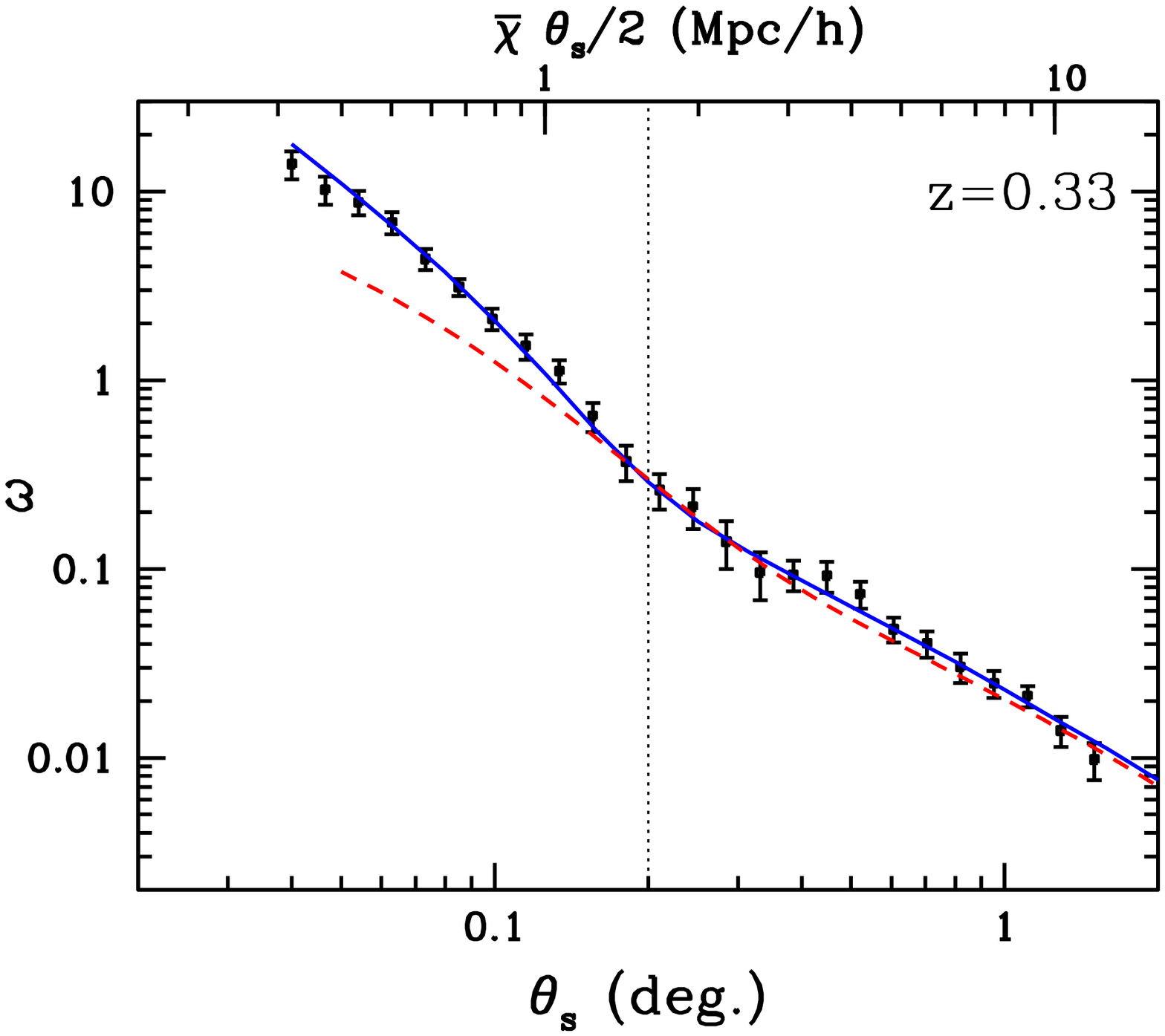}
\includegraphics[width=2.0in]{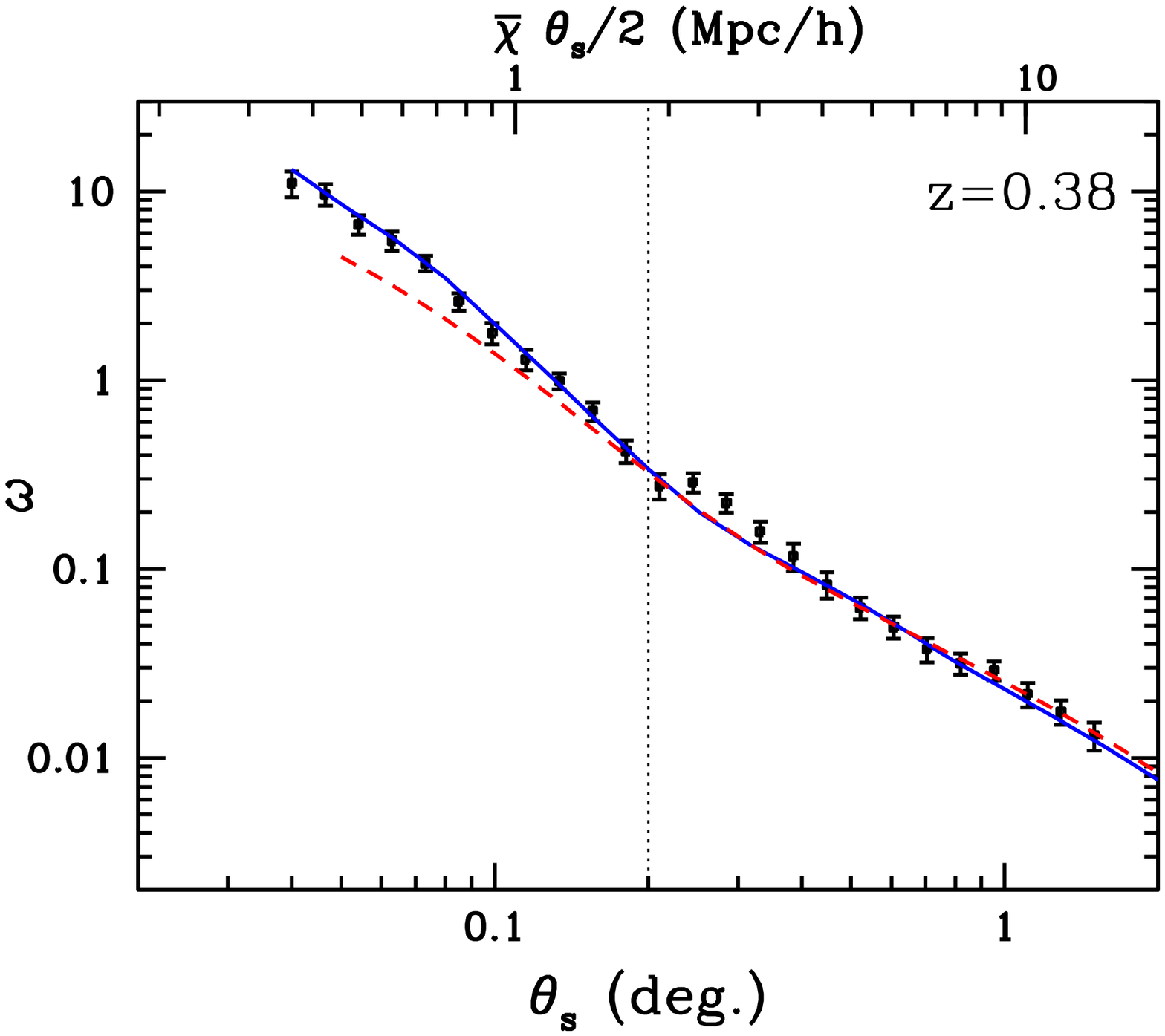}
\includegraphics[width=2.0in]{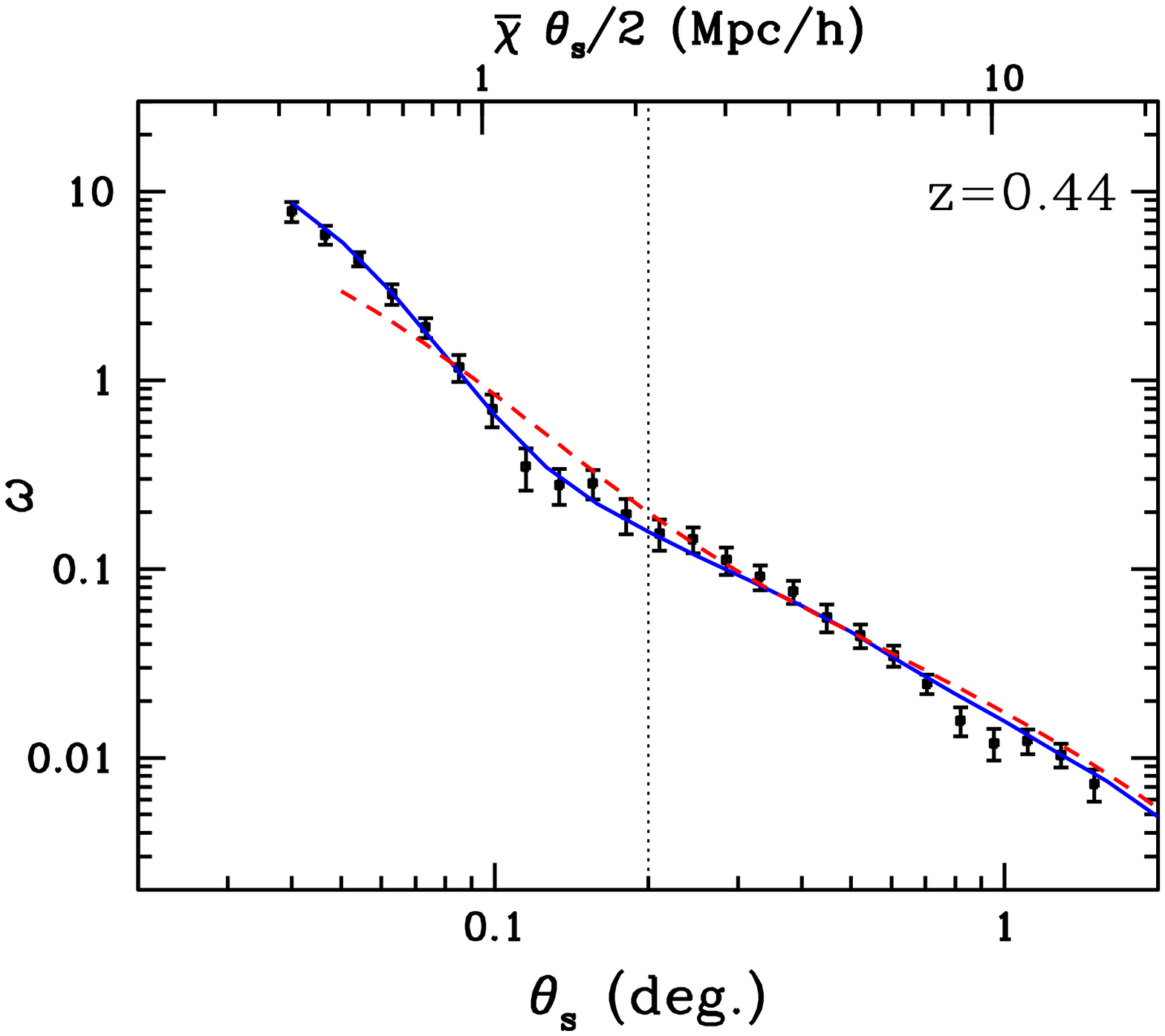}
\includegraphics[width=2.0in]{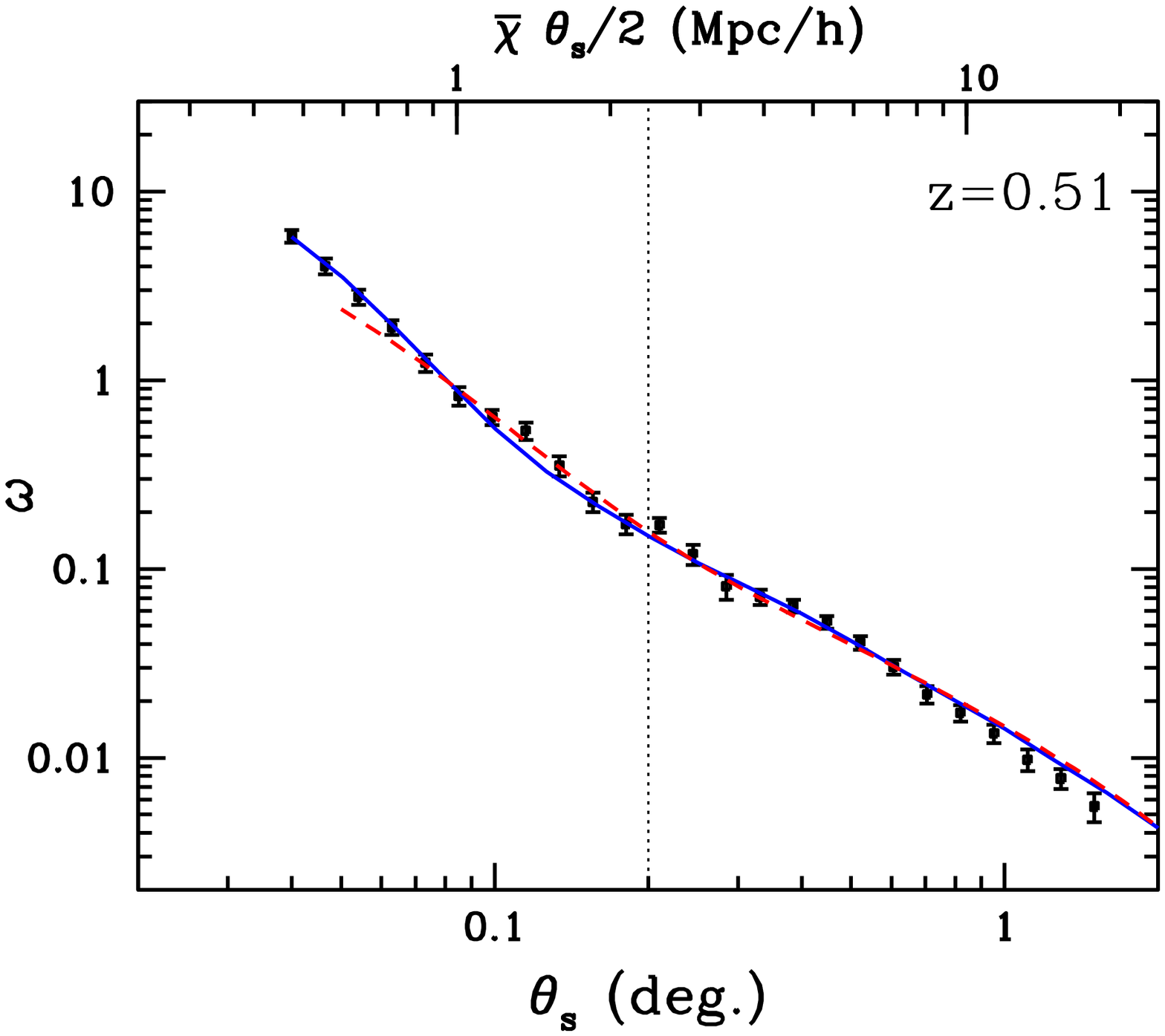}
\includegraphics[width=2.0in]{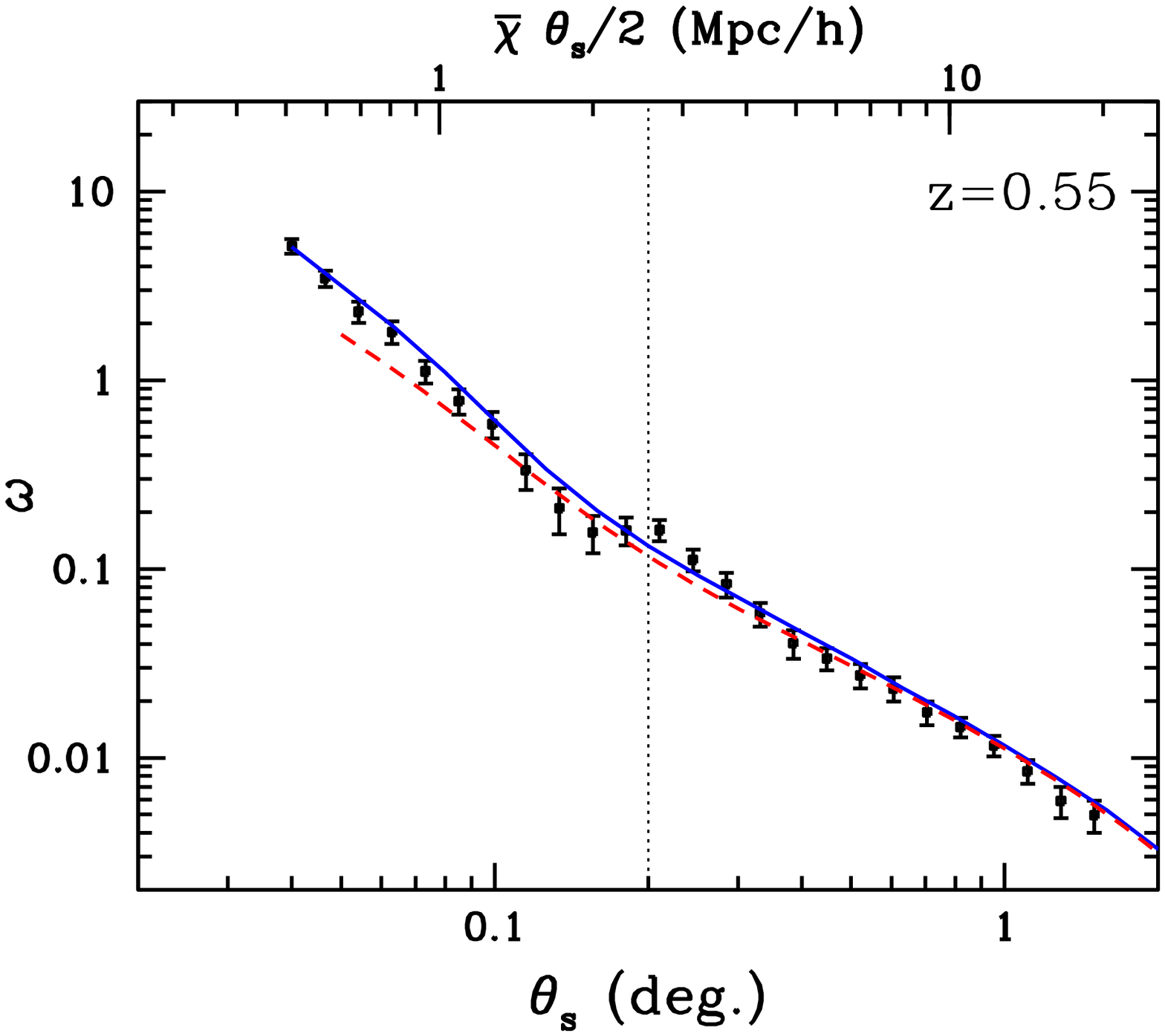}
\end{center}
\caption{The observed $\omega$ for the six LRG redshift slices, as a function
of the filter scale $\theta_s$.  Recall that $\omega$ probes the angular
correlation function on scales $\sim\theta_s/2$; the corresponding physical
scales are also shown.  We also plot the best fit models, both from the
\protect\citet{HaloFit} fitting formula [dashed/red] for the nonlinear dark
matter clustering, as well as from our halo model fits [solid/blue].
The dotted vertical line marks the angular scale beyond which we fit the
{\texttt HaloFit} models.
These fits deviate from the observed clustering on small scales; LRGs are not
distributed like the dark matter on these scales.
The halo model correlation functions are estimated from the same realizations
in the $500\,h^{-1}$Mpc simulation that we use to interpret the quasar-LRG
cross-correlations.  Note that the data are well fit by the halo model on
both small and large scales.}
\label{fig:lrgomega}
\end{figure*}

The angular clustering of the LRGs in the six redshift slices is
shown in Fig.~\ref{fig:lrgomega}, using the estimator described above; also
plotted is the approximate physical scale probed by $\omega$ at a given $\theta_{s}$.
Note that the errors between different points are correlated, and we use the 
full covariance matrix in all fits.

Given the observed angular clustering, one can infer the underlying three
dimensional clustering of the sample; we do this using two methods.
The first assumes that the LRG clustering traces the dark matter with a
scale-independent bias on large scales.  We compute $\omega(\theta_s)$
assuming the \citet{HaloFit} prescription for the shape of the nonlinear
dark matter power spectrum, and estimate the large-scale bias by fitting
to the data on scales larger than $0.2^{\circ}$, corresponding to physical
scales $\approxgt 2\,h^{-1}$Mpc.  The best fit models are plotted in
Fig.~\ref{fig:lrgomega}, with the corresponding bias values in
Table~\ref{tab:lrg}.
The \citet{HaloFit} prescriptions deviate from the observed clustering on
small scales; as one might expect: LRGs do not trace the dark matter on
these scales.

\begin{figure*}
\begin{center}
\leavevmode
\includegraphics[width=5.0in,angle=270]{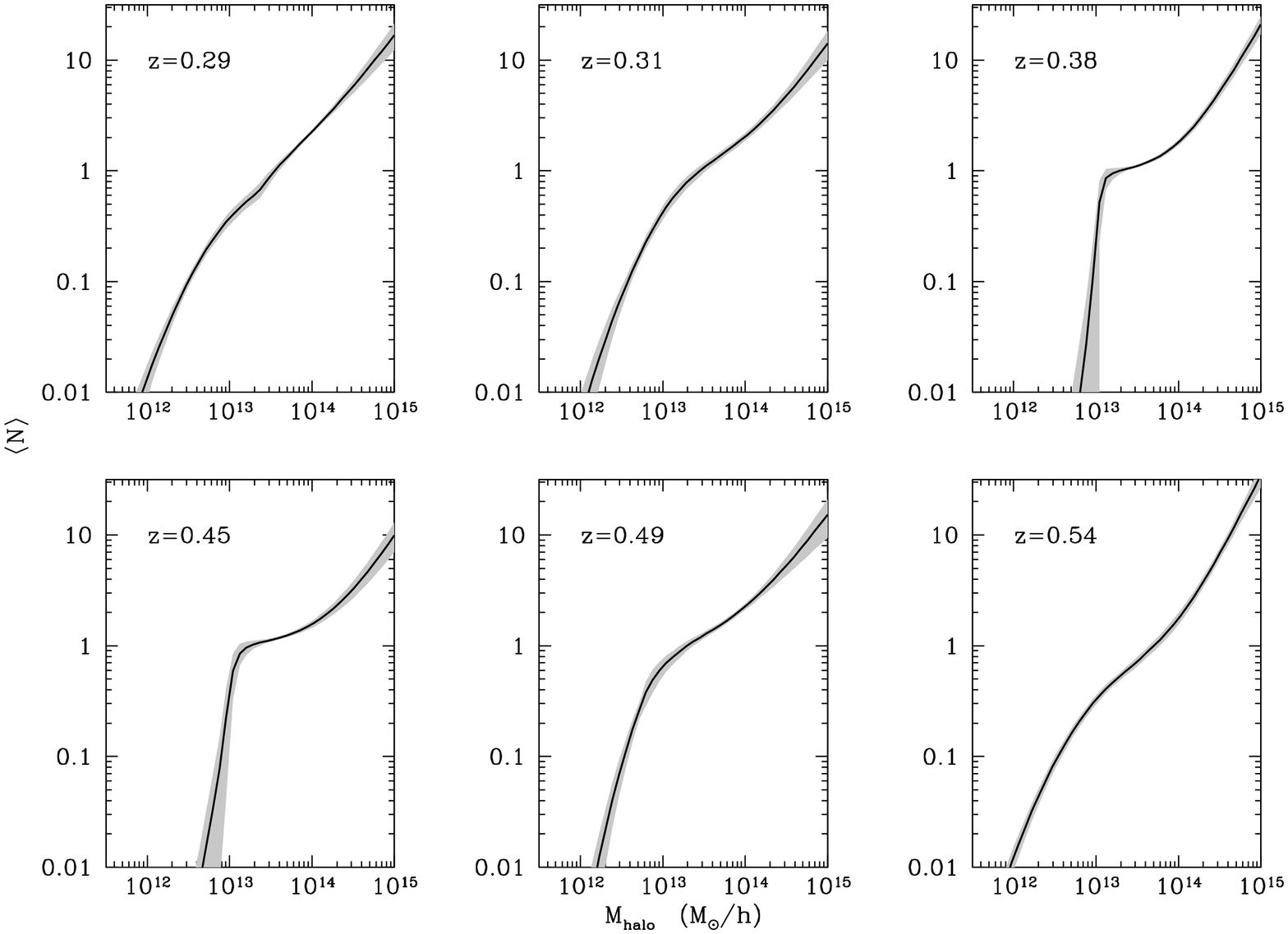}
\end{center}
\caption{Best fit HODs for the six LRG slices we consider in this paper.
The shaded region denotes the errors, as estimated by Monte Carlo.}
\label{fig:hod_500_25_10}
\end{figure*}

The second method attempts to model the observed clustering by populating
halos in a dark matter simulation with LRGs; we refer the reader to 
Appendix \ref{appendix:halomodel} for details.
The best fit $\omega$ are in Fig.~\ref{fig:lrgomega}, and the predicted
large scale bias values are in Table~\ref{tab:lrg}.  Taking into account
the likelihood of scale-dependent bias and the 5 per cent systematic
uncertainty in the halo modeling (see Appendix \ref{appendix:halomodel})
these are in reasonable agreement with the simple fits described above.
The halo models (Fig.~\ref{fig:hod_500_25_10}) reproduce the observed 
correlations well, including the
prominent break in the correlation function (the best fit $\chi^{2}$ values
are in Table~\ref{tab:lrghod}).
Our LRG samples populate a broad range of halo masses, with an approximate
power law dependence of the mean number of LRGs per halo with the halo mass.
Furthermore, we find that halos with masses $\sim 10^{13} M_{\odot}$ have
one LRG in them.
An important by-product of this process is that we obtain mock realizations
of the LRG sample; we use these below to interpret the clustering measurements
of the quasars.

The LRG clustering amplitude is consistent with being non-evolving with 
redshift, implying a bias that evolves as $b(z) \sim 1/D(z)$. An exception
are LRG3 and LRG4, which have a significantly higher bias. The reasons
for this were discussed in detail by \cite{Pad07}.
Comparing the bias values and HOD fits, we see that the LRGs can 
be conveniently grouped into three slices of width $dz_{\rm photo}=0.1$,
each of which samples a homogeneous population of galaxies. 

Finally, we note our results are consistent with those of \cite{Pad07} correcting
for differences in the fiducial cosmologies. That paper also performed
a number of systematic tests on the LRG sample, and we simply refer
the reader to that work instead of repeating them here.

\subsection{QSO Clustering}
\label{sec:qsocluster}

\subsubsection{Methods}

The quasars in our sample have spectroscopic redshifts and thus we know
(up to small uncertainties due to peculiar velocities) a physical distance
to each object.  This allows us to work in terms of transverse separation
rather than angular separation, i.e.~to measure
\begin{equation}
  w_p(R) \equiv \int d\Delta\chi\ \xi\left(\sqrt{R^2+\Delta \chi^2}\right)
\end{equation}
rather than $w(\theta)$ or its generalizations. As with the LRGs, this 
is a real space measurement, avoiding the need to model redshift space
distortions.

We start by considering quasars in a narrow redshift range (a comoving
distance $\chi_0$ away), and correlate them with LRGs with a normalized
radial distribution, $f(\chi)$.
In the flat sky approximation, the angular correlation function is given by,
\begin{equation}
  w(\theta) = \int d\chi\ f(\chi)
  \,\xi\left(\sqrt{\chi_0^2\theta^2 + (\chi-\chi_0)^2}\right) \,\,,
\label{eq:wtheta1}
\end{equation}
where the usual second integral over the quasar redshift distribution has been
eliminated because the quasars have spectroscopic redshifts.
We can now make the usual \citet{Limber} approximation for $\chi_0\theta$
much smaller than the scales over which $f(\chi)$ varies.
This allows us to hold $f(\chi)$ fixed at $f(\chi_0)$ in the integral, yielding
\begin{eqnarray}
  w(\theta) &\simeq& f(\chi_0) \int d\chi
  \ \xi\left(\sqrt{\chi_0^2\theta^2 + (\chi-\chi_0)^2}\right) \\
  &=& f(\chi_0) w_{p}(R) \,\,.
\label{eq:wtheta2}
\end{eqnarray}
Note the second use of the assumption of a peaked integral to re-extend the
limits of the integral to $\pm\infty$.
The second equality just recognizes the integral as the projected correlation
function $w_{p}$ at transverse separation $R=\chi_0\theta$.

To generalize to a broad redshift slice, we start with the standard assumption
that $w_{p}(R)$ does not evolve over the slice. One then has two choices -- the
first is to estimate $w(\theta)$ over narrow redshift slices, estimate 
$w_{p}(R)$ for each of these slices using Eq.~\ref{eq:wtheta2} and then average.
This has the disadvantage that each individual $w_{p}(R)$ measurement is extremely
noisy, and potentially sensitive to noise in $f(\chi)$. 
The other approach (which we adopt) makes use of the fact that $\theta$ in 
Eq.~\ref{eq:wtheta2} is simply a label, and can just as easily be replaced by $R$.
This has a simple interpretation - one computes the angular correlation
function over the broad redshift range 
but replaces the angular separation by the transverse separation $R$
computed assuming that the LRGs are the same redshift as the quasar being correlated
with. We use the estimator,
\begin{equation}
\label{eq:wtheta_DDDR}
w_{\theta}(R) = \frac{QG(R)}{QR(R)} - 1 \,\,,
\end{equation}
where $QG$ and $QR$ are the quasar-galaxy and quasar-random pairs, and our notation
makes explicit that we are binning in physical transverse separation. 
Since we assume that $w_{p}(R)$ is constant over the redshift range, Eq.~\ref{eq:wtheta2}
yields
\begin{equation}
\label{eq:wtheta3}
w_{\theta}(R) = \langle f(\chi) \rangle w_{p}(R) \,\,,
\end{equation}
where the average is done over the quasar redshift distribution. Note that 
this formulation
addresses both the problems of the first implementation. Finally, we point
out that, in Eq.~\ref{eq:wtheta_DDDR}, one only requires the angular selection 
function of the LRGs to estimate $QR$.

Note that one could use the estimator described by Eq.~\ref{eq:omegath1} to
measure the cross-correlations. However, the lower signal-to-noise of the 
quasar-LRG cross correlations eliminates the advantages of the estimator, and
therefore, we choose the simpler estimator. 

Estimating the covariance matrix for our sample is simplified by the low 
areal density of the quasars, making them effectively independent for the 
scales of interest. We therefore estimate the covariance matrix by simply
bootstrapping the individual quasars. 
As with the LRGs, we remove the artificially large condition number by scaling
$w_{p}(R)$ by $R$ before estimating the covariance matrix.
Finally, as a check, we note that we obtain
consistent results if we replace the bootstrap covariance matrix with a jackknife 
estimate.

\subsubsection{Results}
\label{sec:qsoresults}

In order to determine what subsamples to cross-correlate with, we start 
with the observation (see Sec.~\ref{sec:lrgcluster}) that the LRGs can be grouped 
into three homogeneous slices of width $0.1$ in photometric redshift, i.e. 
grouping slices (1,2), (3,4), and (5,6) of Table~\ref{tab:lrg} together.
We use the mean redshift of each LRG redshift slice and its 
width to determine the redshift range of the quasars 
to cross-correlate with;
this defines the redshift ranges of the quasar subsamples in Table~\ref{tab:summary}.
Note that since we use the true redshift distribution of the LRGs (as opposed to 
the photometric redshift distribution) to determine the mean and width of the slices,
we automatically correct for any biases and asymmetries in the photometric 
redshift errors. For each of these three quasar redshift slices (denoted as ALL below), 
we further consider the following subsamples - the LSTAR sample defined in 
Sec.~\ref{sec:data}, restricting to quasars targeted with the latest version
of the SDSS target selection algorithm, and a bright and faint subsample
split at the median luminosity for each redshift slice. 

\begin{figure*}
\begin{center}
\leavevmode
\includegraphics[width=2.0in]{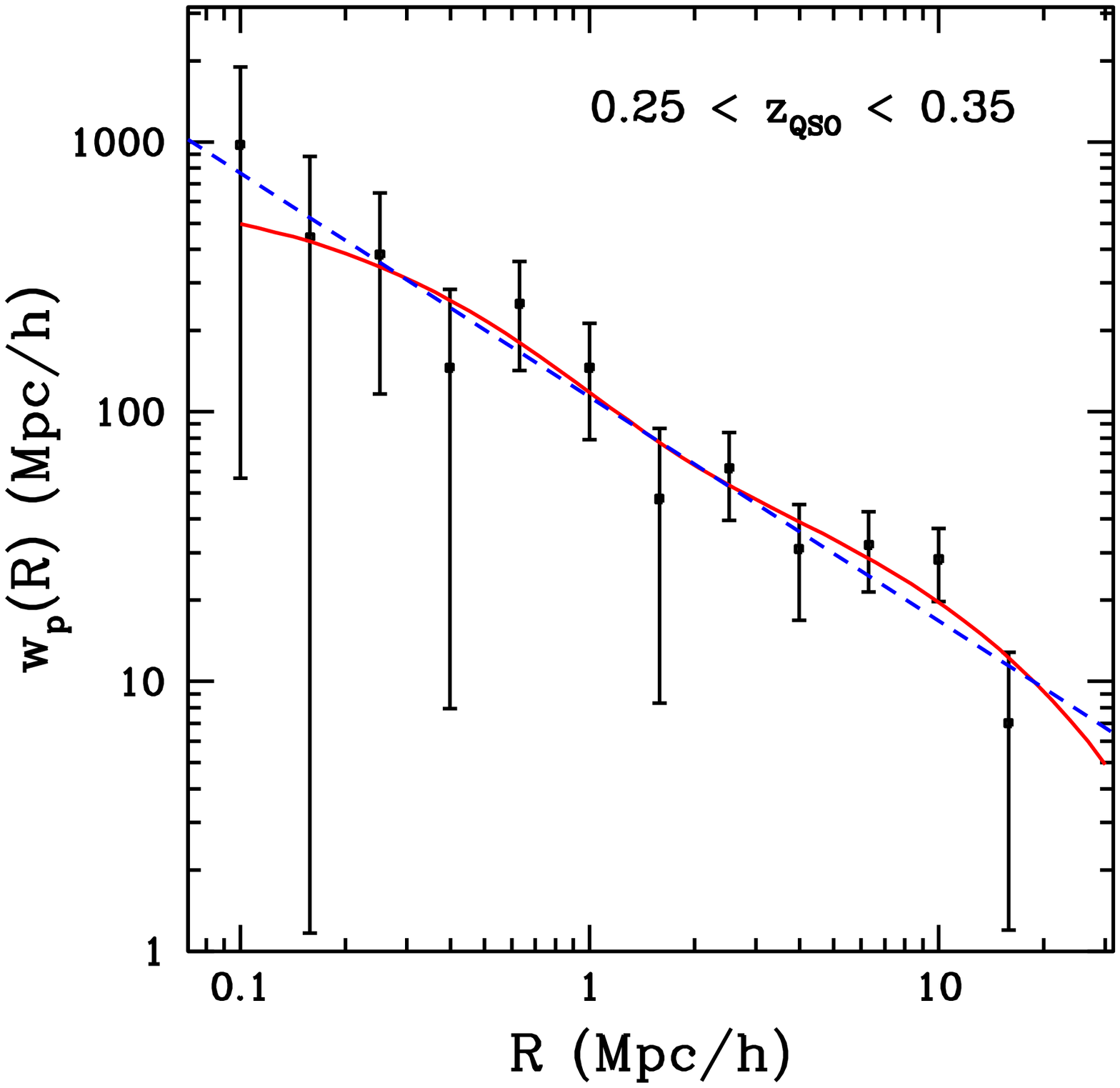}
\includegraphics[width=2.0in]{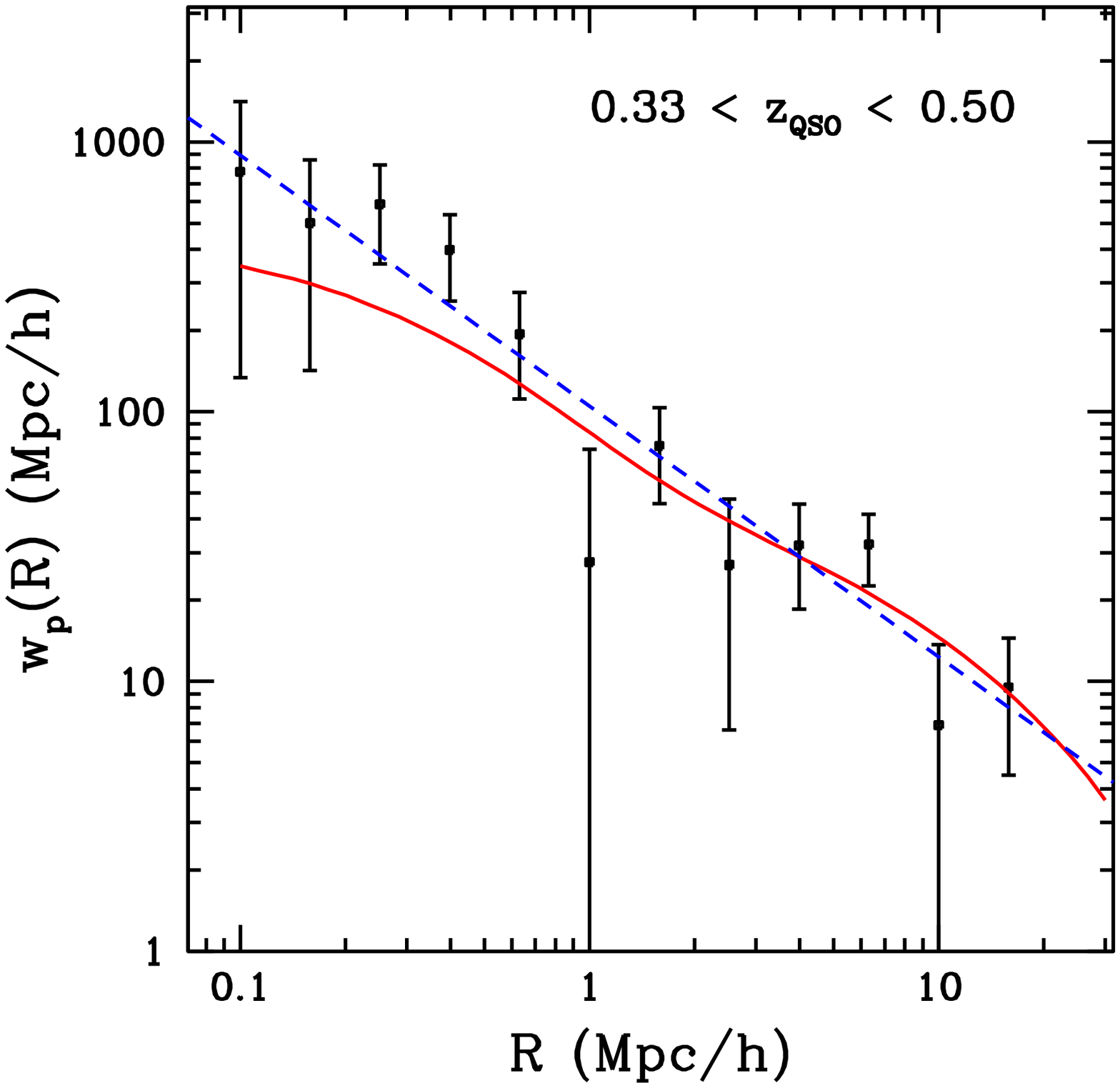}
\includegraphics[width=2.0in]{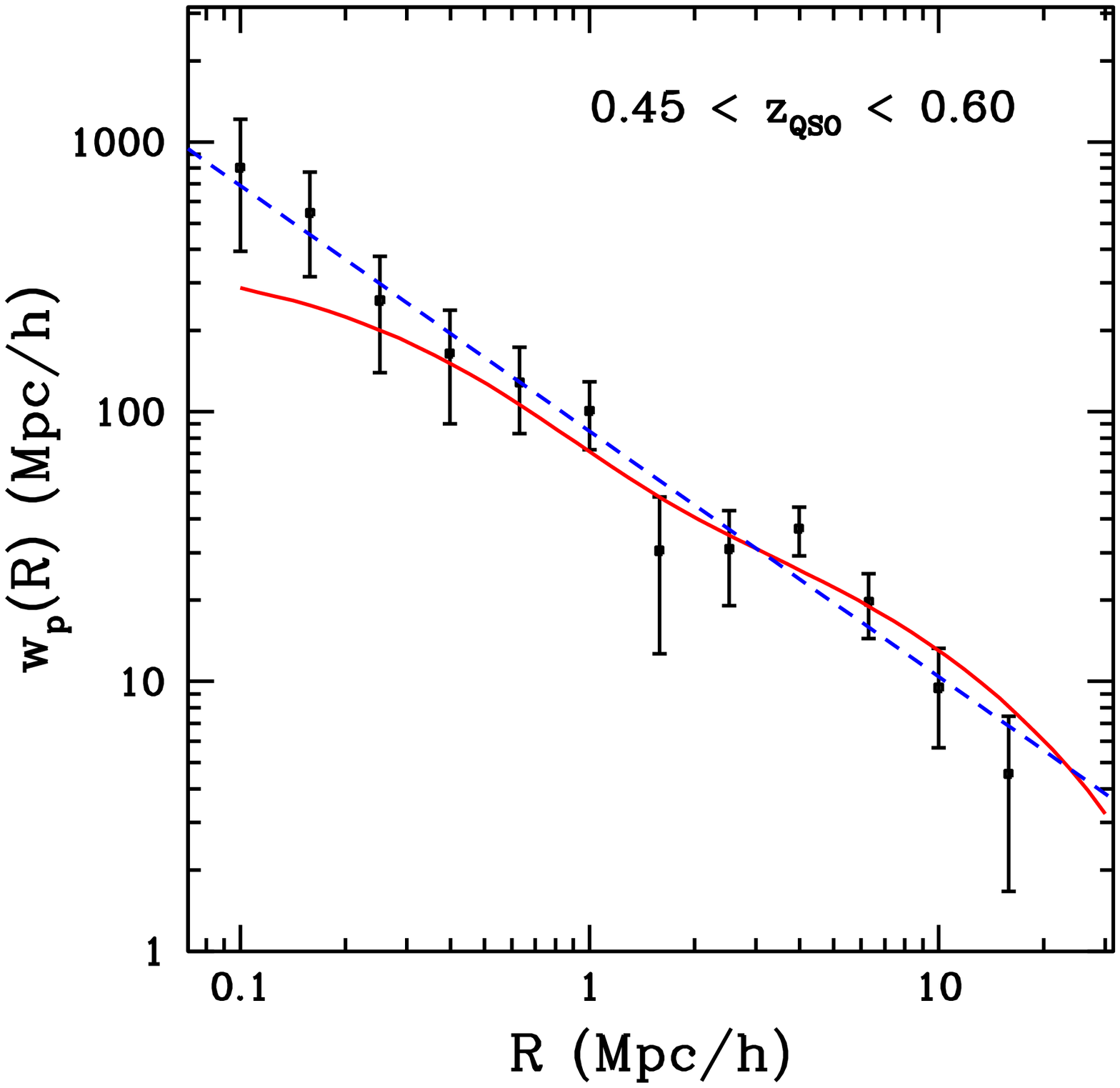}
\end{center}
\caption{The observed $w_{p}(R)$ measuring the cross-correlation between quasars and the LRG samples. 
The solid [red] line shows the {\texttt HaloFit} prescription for the nonlinear dark matter 
correlation function, normalized by a scale independent bias to best fit the observed correlations. 
The dashed [blue] line shows the best fit power law model.
}
\label{fig:qsowp}
\end{figure*}

As anticipated earlier, we find that restricting to the latest version 
of the SDSS target selection algorithm gives the 
same cross-clustering power against the LRGs as the ALL sample.  Since it has better statistics
we will use the ALL sample below. Furthermore, as is evident from Fig.~\ref{fig:zL},
the luminosity baseline is rather small, and no clear trend with luminosity 
emerges, given our errors. 
We therefore present detailed results
for only the ALL and LSTAR samples below.
The higher redshift slices do not extend
significantly below $L_{\star}$, so to simplify the interpretation we use the LSTAR
sample as our fiducial sample. Fig.~\ref{fig:qsowp} plots the cross-correlations
for the LSTAR sample.

We present both power law and large scale bias fits to both 
these subsamples. In order to estimate the mean and error for any parameter
${\bf p}$ (possibly a vector), we use
\begin{equation}
\label{eq:avp}
\langle {\bf p} \rangle = \frac{\int d{\bf p} {\cal L({\bf p})} {\bf p}}
{\int d{\bf p} {\cal L({\bf p})}} \,\,,
\end{equation}
and 
\begin{equation}
\label{eq:errp}
\sigma_{p_{i}}^{2} =  \langle {p_{i}^2} \rangle - \langle {p_{i}} \rangle^{2}
\end{equation}
where the likelihood is defined by
${\cal L}\equiv \exp\left({-\chi^2/2}\right)$,
with $\chi^{2}$ computed using the full covariance matrix.
For the power law fits, we adopt a two parameter model,
\begin{equation}
\label{eq:powerlaw}
\frac{w_{p}(R)}{R} = \frac{\sqrt{\pi}\,\Gamma[(\gamma-1)/2]}{\Gamma[\gamma/2]} 
\left(\frac{r_0}{R}\right)^{\gamma} \,\,,
\end{equation}
which corresponds to a 3D cross-correlation of the form $\xi(r) = (r/r_0)^{-\gamma}$.
We fit this model to the measured correlations on all scales.
In order to determine the large-scale bias we compute $w_p(R)$ for the dark
matter at $z=0.3$, $0.42$ and $0.53$ using the prescription in \citet{HaloFit}.
This is then scaled by an $R$-independent multiplier to obtain the best fit
to the data in the range $R>2\,h^{-1}$Mpc.
In the limit of scale-independent, deterministic bias the multiplier is
$b_Qb_{LRG}$ (see Table \ref{tab:qsobias} for the values).
We transform these into $b_{Q}$ by using $b_{LRG}$ for slices
1, 3 and 5, scaled to the corresponding redshift by the growth factor.
Note that the LRG clustering amplitude is close to constant with redshift,
so we would obtain consistent numbers if we had used the other slices.
The lower limit in the fit, $2\,h^{-1}$Mpc, was determined by the scale
at which $w_p(R)$ from QSO-LRG cross-correlations in our mock catalogs
(see \S\ref{sec:smallscale})
showed significant scale-dependent bias.

The results for the large scale bias and power law fits are in
Table~\ref{tab:qsobias}, while Fig.~\ref{fig:bias_halom} plots the
evolution of the clustering amplitude of the LSTAR sample as a function
of redshift.
Our results are consistent with a constant clustering amplitude from
$z=0.25$ to $0.6$, corresponding to a bias of $1.09\pm0.15$ at $z=0.43$.

Table~\ref{tab:prevwork} summarizes our results compared with previously
published work.\footnote{We do caution the reader that the errors for a
number of these measurements are simply Poisson errors, and ignore correlations
between different scales, and are therefore likely underestimated.}
Our results strongly favour the general consensus that the bias of low
redshift quasars is $\sim 1$; this is also consistent with the 
models of \citet{Hop07} as well as the previous extrapolations by \citet{Cro05}. 
There are two significant exceptions -
\citet{Mye07a} find $1.93\pm0.14$ based on a photometrically selected sample 
of quasars. It is possible that contamination by a high redshift population could boost
the measured bias values. The more intriguing discrepancy is with \citet{Mou08} who analyze a similar
sample to ours, also in cross-correlation with LRGs, and find biases 
between $1.90\pm0.16$ and $1.45\pm0.11$ depending on the particular LRG and quasar sample they
cross-correlate against. These results are also discrepant
with \citet{Ang08} -- with whom we are consistent --
who analyze the same sample in auto-correlations. Furthermore, 
the scatter in the different subsamples analyzed by \citet{Mou08}
significantly exceeds their quoted errors, suggesting either
a systematic in their analysis, or an underestimate of their errors.
Using the observed scatter between the different subsamples as 
an estimate of the error yields a value consistent with our measurement.
Finally, two results not presented in Table~\ref{tab:prevwork} are
\citet{Ser06} and \citet{Str07}.
There are a number of subtleties with interpreting these results (resulting
in misunderstandings in the literature); we therefore defer a detailed
discussion of these results to Appendix~\ref{appendix:serber}.

\begin{table*}
\begin{tabular}{l|cccc|cccccc}
\hline
Sample & LRG Sample & $b_{QSO}b_{LRG}$ & $b_{QSO}$ & $\chi^{2}$ & 
    $r_0$ (Mpc/$h$) & $\gamma$ & $\langle r_0 \rangle$ (Mpc/$h$)& $\langle \gamma \rangle$ & $r$ & $\chi^{2}$ \\ 
\hline
ALL0 & $1$ & $2.73 \pm 0.45$ & $1.60 \pm 0.26$ & $ 1.78$ & 
    6.70 &   1.71 &   6.48 $\pm$   0.63 &   1.70 $\pm$   0.10 &   0.18 &   8.01\\
ALL1 & $3$ & $2.56 \pm 0.47$ & $1.17 \pm 0.22$ & $ 4.03$ & 
    6.44 &   1.82 &   6.29 $\pm$   0.54 &   1.81 $\pm$   0.09 &  -0.30 &   6.70\\
ALL2 & $5$ & $2.30 \pm 0.38$ & $1.20 \pm 0.19$ & $ 2.86$ & 
    5.56 &   1.82 &   5.46 $\pm$   0.42 &   1.82 $\pm$   0.08 &  -0.28 &   5.94\\
\hline
LSTAR0 & $1$ & $2.50 \pm 0.62$ & $1.46 \pm 0.37$ & $ 3.88$ &
    6.60 &   1.83 &   6.23 $\pm$   0.90 &   1.82 $\pm$   0.15 &  -0.13 &   7.42\\
LSTAR1 & $3$ & $2.12 \pm 0.64$ & $0.96 \pm 0.29$ & $ 4.54$ &
    6.00 &   1.93 &   5.75 $\pm$   0.79 &   1.92 $\pm$   0.14 &  -0.49 &  10.22\\
LSTAR2 & $5$ & $2.08 \pm 0.41$ & $1.08 \pm 0.20$ & $ 5.15$ &
    5.42 &   1.91 &   5.29 $\pm$   0.48 &   1.91 $\pm$   0.09 &  -0.44 &   8.68\\
\hline
\end{tabular}
\caption{\label{tab:qsobias} The large-scale quasar bias and power law fits 
for the ALL and LSTAR
samples. We fit scales $R>2\,h^{-1}$Mpc (5 points) for the bias, and all scales (12 points) for 
the power law model. The second column
lists the LRG slice assumed for the bias (see text and Table~\ref{tab:lrg} for
details), while the third and fourth columns list the amplitude of the
quasar-LRG cross-correlation and the implied (scale-independent) bias, assuming
a dark matter $\xi(r)$ given by the \protect\citet{HaloFit} model.
Note that we ignore the contribution of the error in the LRG bias in the
derived QSO bias. The best fit $r_0$ and $\gamma$ values are listed under $r_0$ and
$\gamma$, while likelihood-averaged values are under $\langle r_0 \rangle$ and $\langle \gamma \rangle$.
Note that the errors in $r_0$ and $\gamma$ are correlated; the cross-correlation coefficient is under
$r$.}
\end{table*}

\begin{table}
\begin{tabular}{crcc}
\hline
$z$ & $L_{\rm min}$ & $b_{QSO}$ & Reference \\
\hline
$0.25 < z < 0.6$ & $L_{\star}$ & $1.09\pm0.15$ & (1) \\
$0.3 < z < 0.68$ & $0.4L_{\star}$& $1.27\pm0.20$ & (2) \\
$0.4 < z < 1.0$ & $0.1L_{\star}$ & $1.93\pm0.14$ & (3) \\
$0.7 < z < 1.4$ & $0.1L_{\star}$ & $1.09\pm 0.29$ & (4) \\
$z \sim 0.6$    & $0.4L_{\star}$ & $1.10\pm 0.20$ & (5) \\
$z \sim 0.6$& $0.4L_{\star}$ & $1.90\pm0.16$ & (6) \\
$z \sim 0.6$& $2.5L_{\star}$ & $1.45\pm0.11$ & (7) \\
$z < 0.3$ & $0.4L_{\star}$ & $0.97 \pm 0.05$ & (8) \\
\hline 
\end{tabular}
\caption{\label{tab:prevwork} A summary of previous low redshift quasar
clustering results, compared with results in this work, scaled to the
cosmology assumed here. 
(1) This work (2) \protect\citet{Cro05} (3) \protect\citet{Mye07a}
(4) \protect\citet{DEEP2} (5) \protect\citet{Ang08}
(6,7) \protect\citet{Mou08} (8) \protect\citet{Cro04}. For the
results from \protect\citet{DEEP2}, we scale the relative bias presented
there by the large scale bias $b=1.22$ of all DEEP2 galaxies
\citep{Zhe07}}
\end{table}

\section{Interpretations}
\label{sec:interp}

\subsection{The Large Scale Bias}
\label{sec:largescale}

\begin{figure}
\begin{center}
\leavevmode
\includegraphics[width=3.0in]{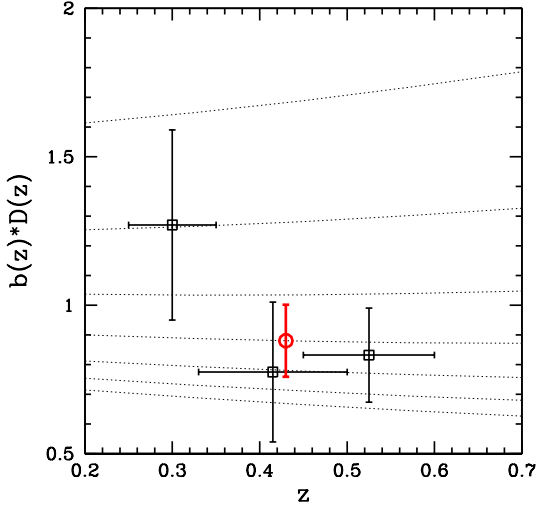}
\end{center}
\caption{The bias of the three LSTAR quasar subsamples [squares] as a
function of redshift, as well as the mean bias [circle].
We consider the bias scaled by the growth factor to focus on the evolution
of the clustering amplitude.
Also plotted [dashed lines] is the bias of halos
with masses (from bottom to top) ranging from $\log_{10}(M/M_{\odot})$ of 10.5
to 13.5 in steps of $d\log_{10}(M/M_{\odot}) = 0.5$.
Note that the mean large scale bias of $1.09\pm0.15$ at $z=0.43$ suggests that
quasars live in halos of mass $\sim 10^{12} h^{-1} M_{\odot}$.}
\label{fig:bias_halom}
\end{figure}

\begin{figure}
\begin{center}
\leavevmode
\includegraphics[width=3.0in]{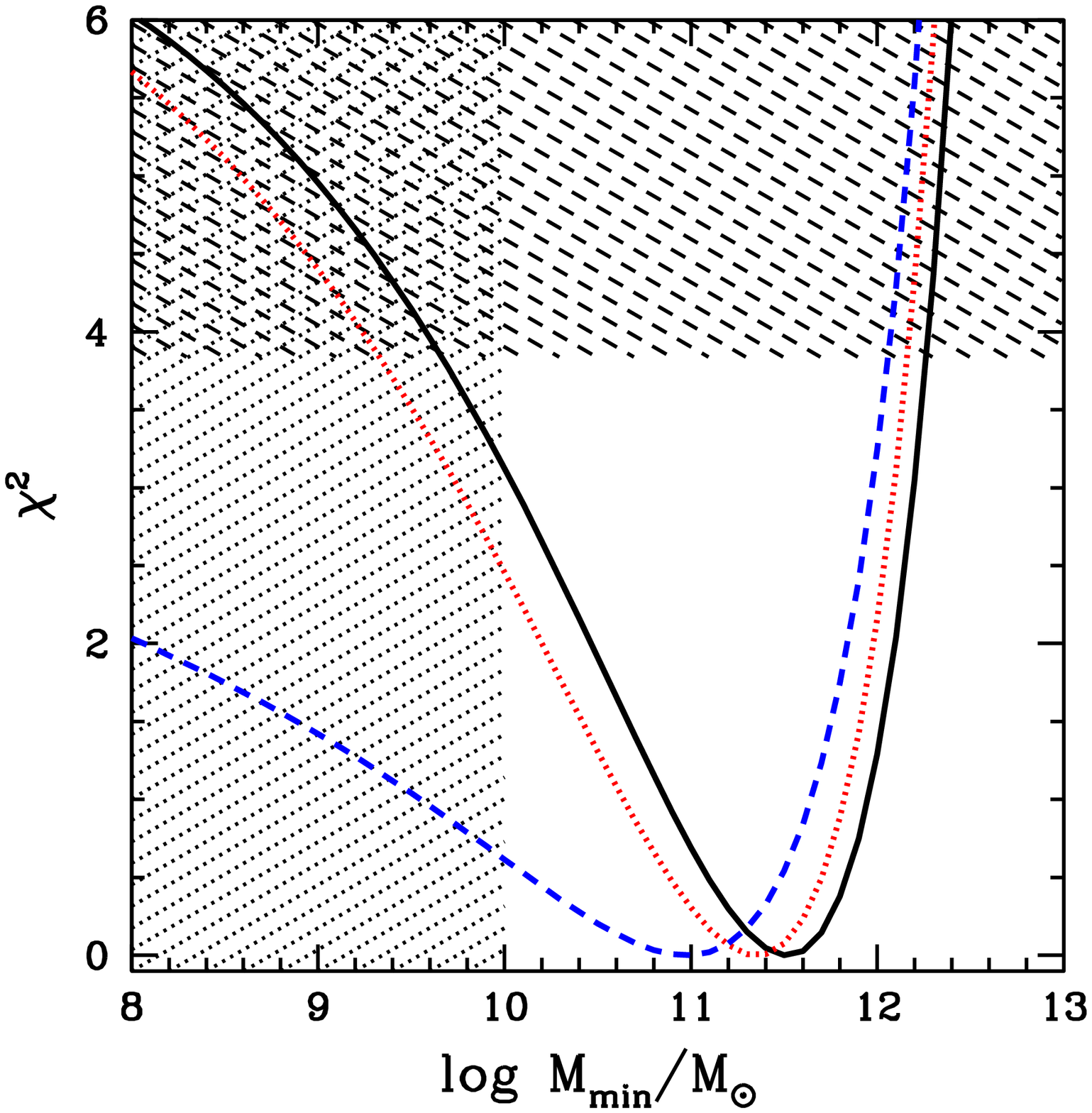}
\end{center}
\caption{$M_{\rm min}$ for the threshold [solid, black] and galaxy-like
$\alpha=0.5$ [dotted,red] and $\alpha=0.9$ [dashed, blue] HODs, as constrained
by the average large scale bias of the LSTAR sample.  The dotted hatched
region marks the region disfavoured by quasar lifetimes and the
$M_{\rm bh}-\sigma$ relation.  The dashed hatched region is excluded at 95
per cent confidence.  See the text for more details.}
\label{fig:hod_bigbias}
\end{figure}

The large-scale bias of any population of objects provides information on
the mean dark matter halos mass hosting that population.  Specifically, if
$N(M_h)$ is the mean number of QSOs hosted by a halo of mass $M_h$ then
\begin{eqnarray}
\bar{n} &=& \int dM_h\ \frac{dn_h}{dM_h} N(M_h)\\
\langle b\rangle &=& \bar{n}^{-1}
  \int dM_h\ \frac{dn_h}{dM_h}b_h(M_h) N(M_h)
\label{eqn:nbcal}
\end{eqnarray}
where $dn_h/dM_h$ is the (comoving) number density of halos per mass interval
\citep[e.g.][]{ST99}
and $b_h(M_h)$ is the bias associated with halos of that mass
\citep[e.g.][]{ColKai89, ST99, SMT01}.
Note that $\langle b\rangle$ is independent of the normalization of $N(M_h)$.
Also, since observable quasar activity
is a transient property, the observed quasar number density $\bar{n}_{QSO}$ 
depends upon the average duty cycle, $f_{\rm on}$,
\begin{equation}
\label{eq:foneq}
\bar{n}_{QSO} = f_{\rm on} \bar{n} \,\,.
\end{equation}

To set the scale, Fig.~\ref{fig:bias_halom} compares the observed
large-scale bias with that of halos of fixed mass in our assumed cosmology.
Our data are consistent with host halos having a mass
$10^{11.5}-10^{12.5}\,h^{-1}M_\odot$, in agreement with earlier work
\citep{PorMagNor04,Cro05,PorNor06,Lid06}. \footnote{A little appreciated uncertainty
in this conversion comes from differences in fitting functions to
$b_{h}(M)$ resulting in an additional error of 50 per cent (0.2 dex) in mass.}
Note that our constraints are significantly stronger for higher as opposed
to lower halo masses, since $b_{h}(M)$ is a rapidly rising function above
$\sim 10^{13}\,h^{-1}M_\odot$ but slowly asymptotes to a constant
($b \sim 0.5$) for lower masses.  As we see no evidence for evolution in the
QSO bias, we use the mean value from all three slices below.

However we don't expect QSOs to inhabit halos of a single mass.  To place
constraints on the range of halos in which quasars may be active we consider
two illustrative models. First we imagine that QSOs brighter than $L_{\star}$ 
live in halos more
massive than $M_{\rm min}$, with each halo above $M_{\rm min}$ hosting
exactly one QSO with probability $f_{\rm on}$.  This gives
$N(M_h)=\Theta(M_h-M_{\rm min})$, where the Heaviside $\Theta$
function is unity for positive arguments and zero otherwise.
Fig.~\ref{fig:hod_bigbias} and Table~\ref{tab:bias_quant} summarize
the constraints on this model. As anticipated by our simple scaling argument
above, our measurements suggest $M_{\rm min}\sim 10^{11}\,h^{-1}M_\odot$
corresponding to an average halo mass
$\langle M\rangle\sim 10^{12}\,h^{-1}M_\odot$.
We further strongly disfavour models with 
$M_{\rm min} \gg 10^{12}\,h^{-1}M_\odot$.
Our lower mass limits are significantly weaker for the reasons discussed
above, requiring $M_{\rm min} > 10^{9.5}\,h^{-1}M_\odot$, with even lower
masses still providing marginally acceptable fits.
However, such masses are disfavoured by the locally observed
$M_{\rm bh}-\sigma$ relation \citep{Fer00,Tre02}.
Assuming an Eddington limited accretion rate, the LSTAR sample should be
powered by black holes with masses $M_{bh} > 10^{7}\,h^{-1}M_\odot$, which
live in bulges with $\sigma \sim 100 {\rm km/s}$ or $M_{\rm bulge}$
a few times $10^{9}\,h^{-1}M_\odot$.  Given mass-to-light ratios of a few,
this disfavours $M_{\rm min}<10^{10}\,h^{-1}M_\odot$.

\begin{table}
\begin{center}
\begin{tabular}{lccc}
\hline 
HOD 
& $M_{\rm min}$ & $\langle M \rangle$ & $M_{\rm min}^{95}$  \\
\hline
Thresh. &  0.3 &  1.9 &  2.0 \\
$\alpha=0.5$ &  0.3 &  3.1 &  1.6 \\
$\alpha=0.9$ &  0.1 &  6.1 &  1.3 \\
\hline
\end{tabular}
\end{center}
\caption{Halo masses derived from the average large-scale bias of our QSO
sample.  The masses are quoted in units of $10^{12}\,h^{-1}M_\odot$. 
The different HOD models are described in detail in
the text.  The superscript $95$
refers to the 95 per cent c.l. upper limits. }
\label{tab:bias_quant}
\end{table}

The second model we consider starts from the assumption that QSOs cluster
like a random sampling of a luminosity or color subsample of galaxies.
This motivates a form,
\begin{equation}
\label{eq:galaxyHOD}
N(M_{h}) = \Theta(M-M_{\rm min}) \left[1 + 
    \left(\frac{M}{20M_{\rm min}}\right)^{\alpha}\right] \,\,,
\end{equation}
found to be a good description of galaxies at both low and high redshifts
\citep[e.g.][]{Zeh05,Con06,Whi07}.
This has a central galaxy in all halos above $M_{\rm min}$ and on average
$(M/20 M_{\rm min})^{\alpha}$ satellites in each halo, where the factor of
20 is inspired by fits to SDSS galaxies and N-body simulations.
We consider cases with $\alpha=0.5$ and $0.9$; the former to model blue
galaxies (which are under-respresented in high-mass halos) and the latter
to describe a luminosity selected sample.
(The models are supposed to be illustrative.)
As before, a fraction $f_{\rm on}$ of the QSOs are ``on'' at the time of the 
observation.
Fig.~\ref{fig:hod_bigbias} and Table~\ref{tab:bias_quant} again summarize the
constraints from the large-scale bias, which are very similar to those obtained
for the threshold models.
Within the context of this model we can translate this into a constraint on
the space density of quasar hosts and hence on their luminosity.
The preferred value of $M_{\rm min}$ suggests
$\bar{n}\simeq 5\times 10^{-2}\,h^3\,{\rm Mpc}^{-3}$ or a luminosity
$L< 0.1\,L_\star$ using the blue-galaxy LF of \citet{Fab07}.
The 95 per cent confidence level upper limit on $b$ gives
$\bar{n}\simeq 4\times 10^{-3}\,h^3\,{\rm Mpc}^{-3}$ or a luminosity
$L\sim L_\star$.
This suggests these low redshift quasars live in relatively faint galaxies.
The number density at fixed bias is lowered if we allow scatter in the $L-M$
relation (i.e.~a smooth turn on in Eq.~\ref{eq:galaxyHOD}), as is likely.
With log-normal scatter in $M$ at fixed $L$ of $\sigma_{\ln M}=1$ the space
density is reduced a factor of $\sim 4$ at fixed bias and the upper limit
on the threshold luminosity becomes $1.1\,L_\star$
\citep[c.f.][]{WhiMarCoh08}. Using the stellar mass functions in 
\citet{Bun07}, we find that these number densities correspond to stellar 
masses $M_{\star} < 2 \times 10^{11} h^{-1} M_\odot$.
Of course QSO hosts may not inhabit halos in the manner assumed by
Eq.~\ref{eq:galaxyHOD}.  If QSO hosts are under-represented in intermediate
mass halos then it is possible to have a lower number density and $b\simeq 1$.
This is what is seen in e.g. a sample of galaxies with 
$M_B<-21$ and a star formation rate $>1\,M_{\odot}\,{\rm yr}^{-1}$ in 
the Millennium simulation \citep{Spr05,Cro06,Luc07}. The HOD of these galaxies
is approximately described by a log-normal distribution peaking at 
$M \sim 10^{12} h^{-1} M_\odot$, with a power law distribution at high masses. There
is however a deficit of galaxies at $\sim 10^{13} h^{-1} M_\odot$; this allows 
one to have a low number density, without a high bias.
Note that while our data are unable to constrain such flexible models, 
a large fraction of this uncertainty derives from the fact that the QSO number
density does not add any constraints to the HOD.  

Given the above models and caveats, and the observed space density of quasars,
we constrain the duty cycle (Eq.~\ref{eq:foneq}) to be $<\mathcal{O}(10^{-3})$,
consistent with estimates by \citet{Dun03} from the luminosity function. 
Converting the duty cycle, $f_{\rm on}$, into a lifetime is somewhat ill
defined.  If we assume $t_Q=f_{\rm on}t_H$, with $t_H$ the Hubble time, we
find $t_Q< 10^7\,$yr. 
These lifetimes are broadly consistent with those derived at $z \sim 2$.
On the other hand, the Hubble time is significantly longer, and the duty cycles are
significantly lower. 

The exact timescale to use in the above conversion is not well defined;
our choice of the Hubble time for the halo lifetime
is an approximation \citep[see e.g.][for a different approximation]{MarWei01}.
The differences between the various choices are of the same
order of magnitude as the systematics in modeling the quasar host number 
density, and we therefore opt for simplicity.
We however caution the reader that these
numbers should only be treated as order-of-magnitude estimates.

If QSOs are radiating at the Eddington limit $L_{\rm edd}$ then the minimum $M_{\rm bh}$ in
our LSTAR sample is $3\times 10^7M_\odot$. This value is consistent
with the estimates from the $M_{\rm bh} - M_{\rm halo}$ relation \citep{Fer02},
$M_{\rm bh} \sim 2 \times 10^{7} - 3 \times 10^{9} M_\odot$, with the differences
coming from different assumptions about the halo profiles. 
This suggests $L/L_{\rm edd} \sim 0.01 - 1$, consistent with the results of 
\citet{Cro05} and \citet{Ang08}, although we find no evidence of super-Eddington
accretion.

\subsection{Small Scale Clustering}
\label{sec:smallscale}

\begin{figure}
\begin{center}
\leavevmode
\includegraphics[width=3.0in]{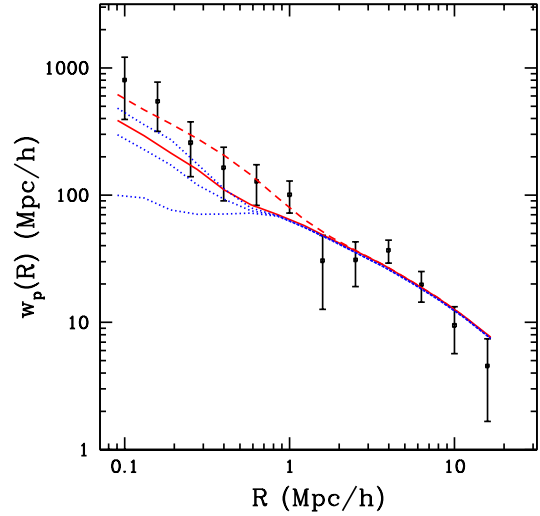}
\end{center}
\caption{Comparison of the cross-correlation of simulated LSTAR2 and LRG5
samples with the observations. The solid ($\chi^2=12.7$) and dashed
($\chi^2=8.1$) lines assume a galaxy-type $N(M_{h})$ with $\alpha=0.3$ and
$0.9$ respectively, while the dotted lines (bottom to top) assume a threshold $N(M_h)$
with a mass-independent satellite fraction of 0,0.5 and 1 
($\chi^2=19.8$, $15.0$ and $11.7$ respectively).  We fix $M_{\rm min}$ to the
best fit values in Table~\ref{tab:bias_quant}; note that the large scale
clustering for all the models are identical.
The $\chi^2$ values are computed using all 12 points and the full
covariance matrix.}
\label{fig:wpmodel}
\end{figure}

Due to the larger number density of LRGs with which we have cross-correlated
our QSOs we are able to measure the clustering down to smaller scales than
would otherwise have been possible.  Interpreting the small-scale clustering
is however complicated by the uncertain relation between the galaxies which
host active QSOs, the galaxies which are selected as LRGs and their parent
host halos.  A full interpretation would require knowledge of the joint
distribution $P(N_{\rm qso},N_{LRG}|M_h)$, which cannot be meaningfully
constrained with the limited data we have available.
It is however straightforward, with the aid of the mock catalogs described
previously, to predict the cross-clustering for any well specified QSO
scenario.  In this section we examine the illustrative models introduced in
the last section, focussing on the $z\simeq 0.5$ data for definiteness.

We begin with the model in which QSOs are hosted exclusively by the central
galaxies of halos above some threshold $M_{\rm min}$.  As with all of the
models we shall ignore the number density constraint by postulating that
a random fraction, $f_{\rm on}$, of the possible hosts are seen as QSOs at
any given time.
If the probability of a host being on is independent of the host properties
the clustering is unchanged and we can use the host population -- with its
better statistics -- to compute $w_p$.
We find that this model cannot simultaneously account for the measured
large-scale bias and the large amplitude of the small-scale correlation
function (see the lowest dotted line in Fig.~\ref{fig:wpmodel}).
To fit the former the mean QSO-weighted halo mass must be low,
but such low mass halos contain only central LRGs, not satellites, so there
are no QSO-LRG pairs with separations $\mathcal{O}(100\,h^{-1}{\rm kpc})$.
We can keep the halo occupancy the same but distribute the QSOs within
the halos as both centrals and satellites.  We assume satellites
follow an NFW profile \citep{NFW} with a concentration of $10$.
-- the precise details don't matter for our purposes, other choices produce
qualitatively similar results and trends.
By making 50 per cent of the QSOs satellites we boost the power on
small scales (Fig.~\ref{fig:wpmodel}), as it is now possible to have QSO-LRG
pairs in the smaller, lower mass halos.

We can also take the form of Eq.~\ref{eq:galaxyHOD}, imagining that a random
sample of galaxies will host QSOs at the time of observation.
We place a central galaxy in all halos above $M_{\rm min}$ and place a
Poisson number of satellites, with mean $(M/20 M_{\rm min})^\alpha$,
distributed like a $c=10$ NFW profile.
The large-scale bias is again set by the mean halo mass, which is larger
for larger $\alpha$ at fixed $M_{\rm min}$.  As $\alpha$ is increased
the spread in small-scale clustering amplitude with $M_{\rm min}$ decreases,
with models lying very close to the data.
A model with $\alpha\simeq 0.9$ (solid line in Fig.~\ref{fig:wpmodel}) 
gives very good fits, with
$\chi^2/{\rm dof}<1$, when $M_{\rm min}\sim 10^{11}\,h^{-1}M_\odot$
but lower values of $\alpha$ (dashed line in Fig.~\ref{fig:wpmodel}) 
are not excluded.
Due to the strong covariance between the $w_p(R)$ points, the constraint
on $M_{\rm min}$ from the full model is not stronger than that from just
the large scale points.

\section{Conclusions}
\label{sec:conclusions}

We measure the small scale clustering of a sample of $\sim~400,000$ photometric
luminous red galaxies and their clustering around a volume limited
sample of $\sim~2000$ $z < 0.6$ low redshift QSOs. By using a new statistical
estimator, we are able to obtain precise measurements of the LRG angular
correlation function, which coupled with their precise and well-characterized 
photometric redshifts, allowed us to constrain how LRGs populate dark matter halos.
We find that LRGs have a clustering 
amplitude that is consistent with not evolving with redshift, and corresponding
to a large scale bias $b\sim 2$ at $z=0.5$. The best fit halo occupation models
suggest that they occupy halos $> 10^{12} h^{-1} M_\odot$, with approximately
one LRG in every $10^{13} h^{-1} M_{\odot}$ halo.
We use these halo occupation distributions to construct mock catalogs of 
LRGs.
Attempting to match the observed cross correlation of LRGs with QSOs by
populating these same mock catalogs with QSOs allowed us, for the first
time, to start to probe how quasars inhabit dark matter halos.

The cross-correlation of QSOs and LRGs is well described on all measured
scales by a power law of slope $\sim 1.8\pm0.1$ and a scale length of
$\sim 6\pm0.5\,h^{-1} {\rm Mpc}$, consistent with observed slopes and
amplitudes for local galaxies.
It is also well described by the nonlinear matter correlation function, scaled
by a constant bias, although there is some evidence for deviations from 
this form at the smallest scales. Such a deviation is however not unexpected,
and is seen for most galaxy samples, which are better described by power laws
down to small scales. Since this is in apparent contradiction with the 
results of \citet{Ser06} and \citet{Str07}, we revisit their measurement
within the framework of cross-correlations developed in this paper 
(Appendix~\ref{appendix:serber}). We explicitly show that their data 
are fit by a power law of slope 1.9, with no deviations on small scales,
and that the claims of an excess come from subtleties in interpreting
their measurements.

The large scale bias $b = 1.09 \pm 0.15$
is consistent with most previous measurements and theoretical models, 
the exceptions being \citet{Mye07a} and \citet{Mou08}; possible reasons for
this discrepancy are discussed in \S\ref{sec:qsoresults}. We see no evidence 
for variations of the bias with redshift or luminosity.
The observed large scale bias constrains quasars to reside in halos
with a mean mass of $10^{12} h^{-1} M_{\odot}$, with uncertainties 
of a factor of a few from the details of how the halos are actually
populated. Our constraints on the halo mass are significantly stronger
from above than below, since these halos are below the nonlinear 
mass scale and occupy the slowly varying region of the halo bias curve.
This should be contrasted with measurements at higher redshifts; even though
the characteristic halo mass is the same at these redshifts \citep{Ang08}, 
it is now higher than the nonlinear mass and probes the steeply rising 
part of the  halo bias curve. This problem is exacerbated when one considers 
realistic models of halo occupation. 

The mean halo mass can, in turn, be used to constrain the
lifetimes of these QSOs \citep{ColKai89,HaiHui01,MarWei01} to be $< 10^{7}\,{\rm yr}$.
This is consistent with measurements at high redshift; on the other hand, the Hubble
time is significantly longer today, and therefore the duty cycles are significantly
shorter.
We also discuss some of the theoretical uncertainties in estimating the number
density of quasar hosts (and therefore the duty cycles and lifetimes), that arise
from the difficulties in constraining how quasars populate dark matter halos. 
We find the number densities of quasar hosts are only certain at the order of 
magnitude level, obviating any need for the detailed modeling (popular in the literature) 
of the conversion of number densities into quasar lifetimes.

Assuming local $M_{\rm bh}-M_{\rm halo}$ relations \citep{Fer02}, we estimate
Eddington ratios between $0.01$ to $1$. We do not find any need for super-Eddington accretion,
in contrast with \citet{Ang08} who require super-Eddington ratios under certain assumptions
for the halo profile. However, the errors on both measurements are large, and are therefore
consistent with each other. These Eddington ratios are also consistent with measurements at $z\sim 2$,
suggesting no evolution in the Eddington ratio with redshift.

Given our detailed modeling of the LRG clustering and the associated mock catalogs, we 
attempt to use the small scale cross-correlation to constrain how quasars must populate
dark matter halos. The size of our errors precludes being able to place 
interesting constraints on general quasar models. However, the forward modeling problem
-- taking a particular quasar model, and comparing it with our data -- is straightforward.
We find that our small scale measurements are inconsistent with quasars being a random subsample of all 
halo centers above a certain mass threshold, but become consistent if we assume $> 25\%$ of
the quasars are satellites. We also find that our data are extremely well fit, if we assume
that quasars are a random subsampling of luminosity-thresholded sample of galaxies, for luminosity 
thresholds between $0.1 L_\star$ to $L_\star$.
The above results suggest that the host galaxies have a number density $< 10^{-3} h^{3} {\rm Mpc}^{-3}$,
corresponding to stellar masses $M_\star < 10^{11} h^{-1} M_\odot$; the exact values
are however sensitive to the particular choice of model.

A second purpose of this paper was to demonstrate the modeling of how 
quasars populate dark matter halos.  An interesting question therefore is how to optimize future
measurements to gain the most leverage on this.
Modeling QSO-galaxy cross-correlations differs from the traditional
galaxy-galaxy auto-correlations in two important ways.
The first is the lack of a constraint on the number density of the underlying
population that quasars are assumed to sample; which usually puts a strong
constraint on the minimum halo mass.
For quasars such a constraint must come from the bias and as discussed above
this is only strongly constraining when one is on the steeply rising part of
the halo bias curve.
The second difference is that the one-halo term is only probed where the QSOs
and galaxies occupy the same halos.  For the quasar and LRG sample presented
here, the mean halo mass for the quasars probes only the tail of the LRG HOD,
making the constraints weaker than one might naively expect.
On the other hand, we emphasize that the accurate and well-characterized
photometric redshifts of the LRGs were an essential prerequisite for doing
the modeling in the first place -- this could not have been done with the
full SDSS
photometric sample.  This suggests that the ideal sample to correlate with 
would be a fainter
sample of red galaxies.
Furthermore, the modeling would be significantly easier
at higher redshifts, where the quasar bias is higher.
Various combinations of these will be available with the next generation of
imaging surveys, making it possible to significantly improve on the
constraints presented here.
 
\medskip

This paper benefited from discussions with Paul Martini and John Silverman.
The authors acknowledge the Aspen Center for Physics for hosting 
a workshop on galaxy clustering in Summer 07, where this work 
was conceived and begun.  NP and MW thank the Center for Cosmology
and AstroParticle Physics at Ohio State University for their hospitality.
NP is supported by a Hubble Fellowship HST.HF-01200.01 awarded by the
Space Telescope Science Institute, which is operated by the Association
of Universities for Research in Astronomy, Inc., for NASA, under contract
NAS 5-26555. 
MW is supported by NASA and the DOE.
PN is supported by a PPARC-STFC PDRA fellowship at the IfA.
This work was supported by the Director, Office of Science, of the U.S. 
Department of Energy under Contract No. DE-AC02-05CH11231.
The simulations used in this paper were performed on the supercomputers
at the National Energy Research Scientific Computing center.

Funding for the SDSS and SDSS-II has been provided by the Alfred P. Sloan
Foundation, the Participating Institutions, the National Science Foundation,
the U.S. Department of Energy, the National Aeronautics and Space
Administration, the Japanese Monbukagakusho, the Max Planck Society, and
the Higher Education Funding Council for England.
The SDSS Web Site is {\tt http://www.sdss.org}.

The SDSS is managed by the Astrophysical Research Consortium for the
Participating Institutions.  The Participating Institutions are the
American Museum of Natural History, Astrophysical Institute Potsdam,
University of Basel, University of Cambridge, Case Western Reserve University,
University of Chicago, Drexel University, Fermilab, the Institute for
Advanced Study, the Japan Participation Group, Johns Hopkins University,
the Joint Institute for Nuclear Astrophysics, the Kavli Institute for
Particle Astrophysics and Cosmology, the Korean Scientist Group,
the Chinese Academy of Sciences (LAMOST), Los Alamos National Laboratory,
the Max-Planck-Institute for Astronomy (MPIA), the Max-Planck-Institute
for Astrophysics (MPA), New Mexico State University, Ohio State University,
University of Pittsburgh, University of Portsmouth, Princeton University,
the United States Naval Observatory, and the University of Washington.

\appendix

\section{Halo model}
\label{appendix:halomodel}

In order to understand the manner in which luminous red galaxies and
quasars inhabit dark matter halos we make use of the halo model
\citep[for a review see, e.g.][]{CooShe02}.
Within this formalism an accurate prediction of galaxy clustering requires
a knowledge of the occupation distribution of objects in halos (the HOD)
and their spatial distribution.
In combination with ingredients from N-body simulations a specified HOD makes
strong predictions about a wide array of galaxy clustering statistics.

Our modeling of galaxy clustering is based on mock catalogs constructed
within the HOD framework by populating halos in a cosmological N-body
simulation.  We use a high resolution simulation of a $\Lambda$CDM cosmology
($\Omega_M=0.25=1-\Omega_\Lambda$, $\Omega_B=0.043$, $h=0.72$, $n_s=0.97$
and $\sigma_8=0.8$).
The linear theory power spectrum was computed by evolution of the coupled
Einstein, fluid and Boltzmann equations using the code described in
\citet{Boltz}.  This code agrees well with {\sl CMBfast\/} \citep{CMBfast},
see e.g.~\citet{SSWZ}.
The simulation employed $1024^3$ particles of mass
$8\times 10^{9}\,h^{-1}M_\odot$ in a periodic cube of side $500\,h^{-1}$Mpc
using a {\sl TreePM\/} code (\citealt{TreePM}; for a comparison with other
N-body codes see \citealt{CodeComparison}).
The Plummer equivalent softening was $18\,h^{-1}$kpc (comoving).
To check for finite volume and force resolution effects we also looked at
simulations of the same cosmology, with the same number of particles, in
boxes $250\,h^{-1}$Mpc and $1\,h^{-1}$Gpc.  The $250\,h^{-1}$Mpc box turned
out to be too small to model LRG clustering.

For each output we generate a catalog of halos using the Friends-of-Friends
(FoF) algorithm \citep{DEFW} with a linking length of $0.168$ times the mean
inter-particle spacing.
This procedure partitions the particles into equivalence classes by linking
together all particles separated by less than a distance $b$, with a density
of roughly $\rho>3/(2\pi b^3)\simeq 100$ times the background density.
A comparison of the mass functions in the $500\,h^{-1}$Mpc and $1\,h^{-1}$Gpc
boxes suggests that at low particle numbers FoF tends to over-count the
number of halos.  To make the mass functions match in the overlap region
we adjusted the mass of the halos downward by a factor $1-n_{\rm part}^{-0.8}$
\citep[for a similar correction see][]{Luk07}.

To make mock catalogs we use a halo model which distinguishes between central
and satellite galaxies. We choose a mean occupancy of halos:
$N(M)\equiv\left\langle N_{\rm gal}(M_{\rm halo})\right\rangle$.
Each halo either hosts a central galaxy or does not, while the number of
satellites is Poisson distributed about a mean $N_{\rm sat}$.
For each sample, we parameterize $N(M)=N_{\rm cen}+N_{\rm sat}$ with 5
parameters
\begin{equation}
  N_{\rm cen}(M) = \frac{1}{2}
  \ {\rm erfc}\left[\frac{\ln(M_{\rm cut}/M)}{\sqrt{2}\sigma}\right]
\label{eqn:ncen}
\end{equation}
and
\begin{equation}
  N_{\rm sat}(M) = \left(\frac{M-\kappa M_{\rm cut}}{M_1}\right)^\alpha
\label{eqn:nsat}
\end{equation}
for $M>\kappa M_{\rm cut}$ and zero otherwise.  Different functional forms
have been proposed in the literature, but the current form is flexible enough
for our purposes.

Central galaxies always live at the minimum of the halo potential while
satellite galaxies are randomly placed assuming an NFW profile \citep{NFW}.
If we instead use randomly chosen dark matter particles within halos,
which preserves the anisotropy of the halos and any substructure, $\xi(r)$
is altered at the 10 per cent level on Mpc scales.  The differences
on large scales are very small.
The concentration of the halo is taken from the N-body simulation, but
multiplied by a free (mass independent) factor to allow galaxies to be
more or less concentrated than the dark matter.  This detail only affects
the predictions on small scales.

Given the 3D galaxy positions the correlation function is computed out
to $10\,h^{-1}$Mpc by direct pair counts, and then extrapolated assuming
constant bias and a (dimensionless) mass power spectrum given by the
$Q$-model
\begin{equation}
  \Delta_m^2(k) = \Delta_{\rm lin}^2(k)\ \frac{1+Qk^2}{1+Ak}
\end{equation}
with $Q=10\,[2/(1+z)]^{0.75}$ and $A=1.7\,h^{-1}$Mpc.
This form provides a reasonable fit to the DM power spectrum in the simulation
over the redshift range of interest.  The correlation function is then
integrated, making the Limber approximation, to find $\omega(\theta_s)$.

Comparison of different paramerizations for $N_{\rm cen}$ and $N_{\rm sat}$,
different methods for making mock catalogs, different techniques for computing
$\omega(\theta_s)$, different ranges and subsets of the data and the different
simulations of the same cosmology, indicate that our results for the
large-scale bias have systematic uncertainties at the several percent level.
In the $500\,h^{-1}$Mpc box different realizations of the same HOD cause 2
per cent changes in the inferred large-scale bias, because of fluctuations in
the galaxy-weighted mean halo mass from Poisson fluctuations in $N_{\rm sat}$.
There is negligible difference in the $1\,h^{-1}$Gpc box.
By running a sequence of boxes of increasing resolution, but with the same
large-scale phases, we find that the cumulative mass function changes by
5-10 per cent in the mass range of interest due to structural changes in the
simulated halos with increasing force resolution.  We might expect a similar
change if we included baryonic cooling and star formation in our simulations.
This affects the inferred number density for a given set of HOD parameters.
However the positions, and hence clustering properites, of the halos are
largely unaffected by increasing force resolution.
In contrast, the bias of a given halo population is sensitive to finite box
size effects on scales a few percent of the box size, and the sensitivity is
larger the more biased the halo population under consideration.
We choose to measure $\xi(r)$ in the $500\,h^{-1}$Mpc box only out to
$10\,h^{-1}$Mpc because we find systematic differences in halo clustering
between the large and small boxes for the rarer halos.  This can be traced
to the particular modes chosen in the initial conditions.  If we restrict to
$10\,h^{-1}$Mpc the large-scale bias agrees between the two simulations to
2 per cent, less than the random error from the fits.
The average halo parameters also agree to within the chain-inferred
dispersion.
Unfortunately the halo bias is still slightly scale dependent at
$10\,h^{-1}$Mpc, as determined from our $1\,h^{-1}$Gpc simulation,
so our results extrapolated assuming constant bias tend to over-estimate
$b$ by 5-10 per cent.  We correct for this over-estimate for the values quoted
in Table \ref{tab:lrg}.

An investigation of all of these effects leads us to assign a 5 per cent
systematic error bar to the large-scale LRG bias estimates we derive.
This uncertainty, while comparable to or larger than the statistical error
for the LRG sample, is irrelevant for our main conclusions.  However, future
work on modeling LRGs for galaxy formation and evolution will be limited by
theoretical uncertainties, and not observational errors.
For making mock catalogs, including QSOs, we use the $500\,h^{-1}$Mpc box.
This allows us to probe further down the mass function.  The theoretical
inaccuracies, of concern for the LRGs, are much smaller than the observational
errors on the QSO-LRG cross-correlations.
As shown in Figure \ref{fig:lrgomega}, the mock catalogs produced in the
$500\,h^{-1}$Mpc box provide a very reasonable description of the LRG
clustering on large and small scales.

\begin{table}
\begin{center}
\begin{tabular}{ccccccc}
\hline
Slice & $M_{\rm cut}$ & $M_1$ & $\sigma$ &
 $\kappa$ & $\alpha$ & $\chi^2$ \\ \hline
 1 & 13.15 & 13.71 &  1.16 &  1.59 &  0.81 & 37.62 \\
 2 & 13.20 & 14.00 &  1.10 &  0.37 &  1.07 & 26.75 \\
 3 & 13.06 & 14.03 &  0.13 &  1.19 &  1.38 & 32.80 \\
 4 & 12.95 & 14.05 &  0.17 &  0.64 &  1.13 & 33.25 \\
 5 & 12.98 & 13.86 &  0.85 &  1.53 &  0.96 & 41.45 \\
 6 & 13.21 & 13.90 &  1.26 &  1.18 &  1.33 & 62.41 \\
\hline
\end{tabular}
\end{center}
\caption{The HOD parameters from our best fitting model and used to make
the mock LRG catalogs.  The meaning of the parameters is given in the text.
The $\chi^{2}$ are for 25 data points.}
\label{tab:lrghod}
\end{table}

\section{Counts of neighbours}
\label{appendix:serber}

Previous authors \citep{Ser06,Str07} estimated the clustering of quasars 
by measuring the overdensities of photometric galaxies around quasars in cylindrical 
apertures. While these measurements can be related to the correlation functions
presented in this paper, there are a number of subtleties in their 
interpretation that have resulted in confusion in the literature. This appendix
attempts to clarify these measurements, as well as compare them with our results.

Following \cite{Ser06}, we define $N_{q,g,r}(R)$ as the average number of photometric 
galaxies within cylindrical apertures of transverse physical (as opposed to angular)
radii $R$, centered on quasars ($q$), spectroscopic galaxies ($g$) and random points ($r$).
If we assume the photometric sample has a normalized redshift distribution $f(\chi)$, 
we use the formalism of \S\ref{sec:clustering}, to write
\begin{equation}
N_{r}(R) = n_{r} \bar{n}_{p} \int_{0}^{R} 2\pi R' dR'
\end{equation}
and 
\begin{eqnarray}
    N_{q,g}(R) = n_{q,g} \bar{n}_{p} \int_{0}^{R} 2\pi R' dR' \times \\
        \left[1 + \langle f(\chi) \rangle_{q,g} w_{qp, gp}(R')\right] \,\,,
\end{eqnarray}    
where $n_{r,q,g}$ is the number of random points, quasars and spectroscopic galaxies,
$\bar{n}_{p}$ is the areal density of the photometric sample, and $w$ is the
projected cross-correlation between the different samples. As in \S\ref{sec:clustering},
$\langle f(\chi) \rangle$ implies the redshift distribution of the photometric sample
averaged over the redshift distribution of the spectroscopic sample.
Normalizing the random galaxies to the number of spectroscopic targets, we then obtain
\begin{equation}
  \frac{N_q}{N_r} = 1 + \frac{\langle f(\chi) \rangle_{q}}{\pi R^2}
  \int_0^R 2\pi R'dR'\ w_{qp}(R')
\label{eq:serberdensity}
\end{equation}
for the quasars with a similar expression for galaxies. We therefore see that the 
overdensities measured in \cite{Ser06} and \cite{Str07} can be related to the 
area-averaged projected correlation function weighted by the redshift distribution
of the photometric galaxies at the redshifts of the spectroscopic targets. This implies
that the overdensities of the quasars and galaxies cannot be directly compared 
since they are scaled by {\it different} weights. 
Furthermore, the 
overall amplitude of the overdensities cannot be interpreted without knowledge 
of the redshift distribution of the photometric galaxies (which is non-trivial
for high redshift quasars since one is starting to probe the high redshift tail
of the photometric galaxies).

Eq.~\ref{eq:serberdensity} has the property that it asymptotes to 1 on large scales
irrespective of galaxy type. We estimate the scale at which $N_{q,g}/N_{r} \sim 1$ as
follows -- the redshift distribution of the photometric
galaxies spans approximately
$1\,h^{-1}$Gpc and the integral of $f$ must equal unity, suggesting $\langle f(\chi) \rangle
\sim 10^{-3} \,h\,{\rm Mpc}^{-1}$. Assuming that the cross-correlation
between quasars (spectroscopic galaxies) and photometric galaxies is similar to 
the quasar-LRG cross-correlation, this implies that the second term of 
Eq.~\ref{eq:serberdensity} is ${\cal O}(1)$ on scales $R \sim 0.1 h^{-1} {\rm Mpc}$. 
Note that this implies that the overdensities are ${\cal O}(1)$ on all scales,
especially on scales larger than a Mpc.
We emphasize that this is simply due to the division by the mean density and in no way implies 
that the quasars and spectroscopic galaxies have the same large scale bias
or inhabit halos of similar masses. 
Note that this is a significant difference between spectroscopic and photometric samples -
for spectroscopic samples, the width of the redshift distribution is typically a 
few tens of Mpc (to integrate out redshift space distortions), and the second term 
in Eq.~\ref{eq:serberdensity} is much larger than the first. 
In this case, the overdensities can be directly
interpreted as the angle-averaged correlation function. For photometric samples, 
the complications can be simply circumvented by subtracting 1 from the overdensities,
if the redshift distribution of the photometric sample is known (see the discussion above).
However, neither \cite{Ser06} nor \cite{Str07} do this when comparing with the
galaxy samples, and their results must not be interpreted as quasars having the same
clustering. An important corollary to this is that the upturn seen in overdensities
cannot be interpreted (as has often been in the literature) as an excess in small-scale
clustering, but is simply the signature of a clustered sample of objects.
Indeed, as we show below, the cross-correlation for the quasars is consistent with
being a power-law down to small scales.

\begin{figure}
\begin{center}
\leavevmode
\includegraphics[width=3.0in]{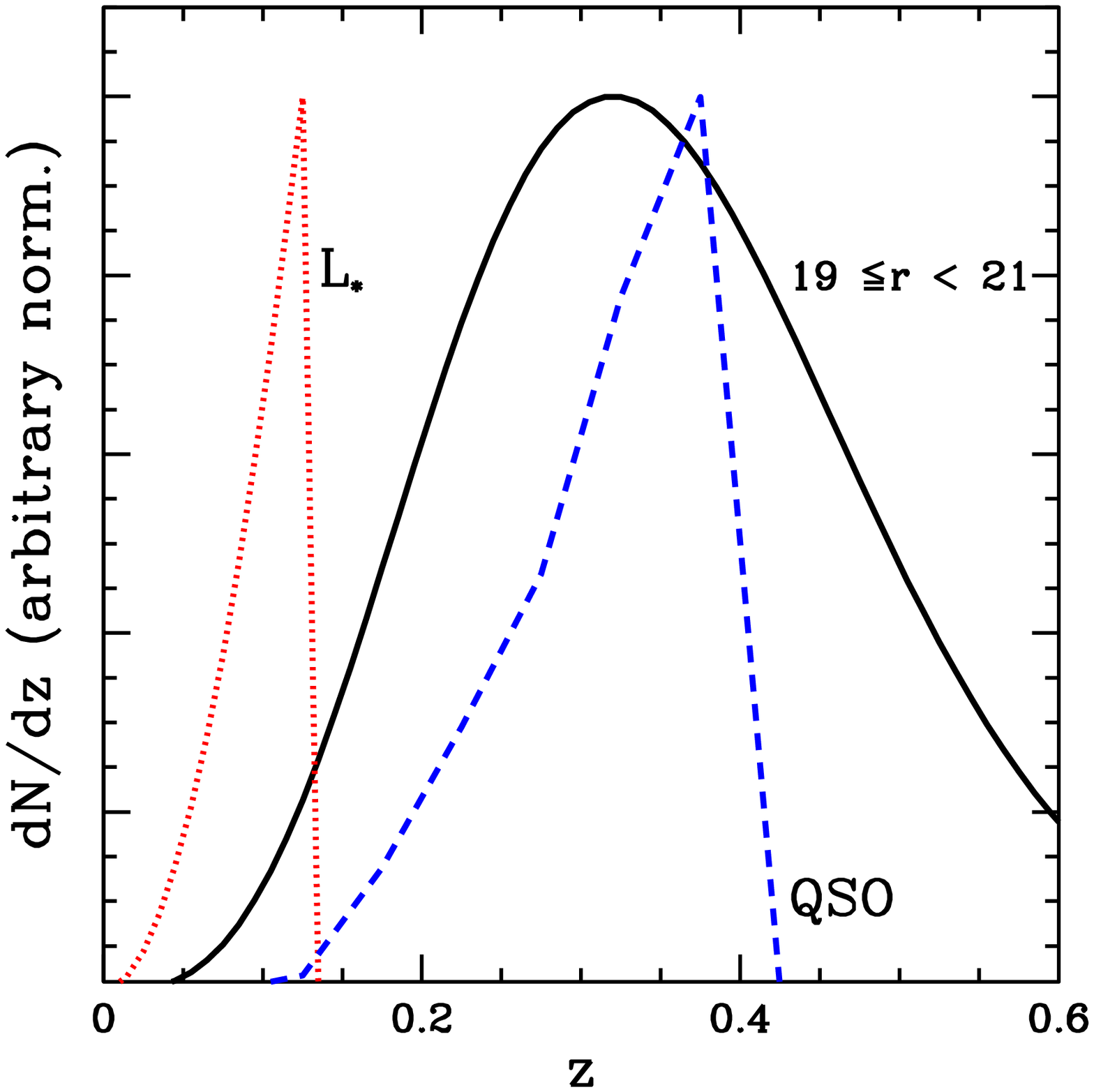}
\end{center}
\caption{The redshift distributions of spectroscopic $L_{\star}$ galaxies [dotted, red], 
photometric galaxies $19 \le r < 21$, [solid, black] \citep{Man07}
and spectroscopic quasars with $z \le 0.4$ [dashed, blue]. Note that the normalizations
of the redshift distributions are arbitrary. The figure emphasizes the fact that 
$L_{\star}$ galaxies and quasars probe very different regions of the photometric redshift
distribution, complicating the comparison between their clustering.
}
\label{fig:lowzdist}
\end{figure}

The final complication in interpreting these results arises due to the width of 
redshift distribution of the photometric sample. Since the photometric sample
covers a large redshift range, it cannot be treated as a homogeneous sample 
and we cannot model its auto-correlation as we
did with the LRGs. In order to interpret the cross-correlation of the quasars
with the photometric sample, we must therefore compare it with the cross-correlation
between a particular population of galaxies and the photometric sample. Unfortunately,
for the SDSS, quasars and galaxies occupy different redshift ranges (see Fig.~\ref{fig:lowzdist}) 
and therefore probe
different sub-populations of the photometric sample. Any comparison 
between quasars and galaxies must also take into account these population 
differences, complicating any analysis.

\begin{figure}
\begin{center}
\leavevmode
\includegraphics[width=3.0in]{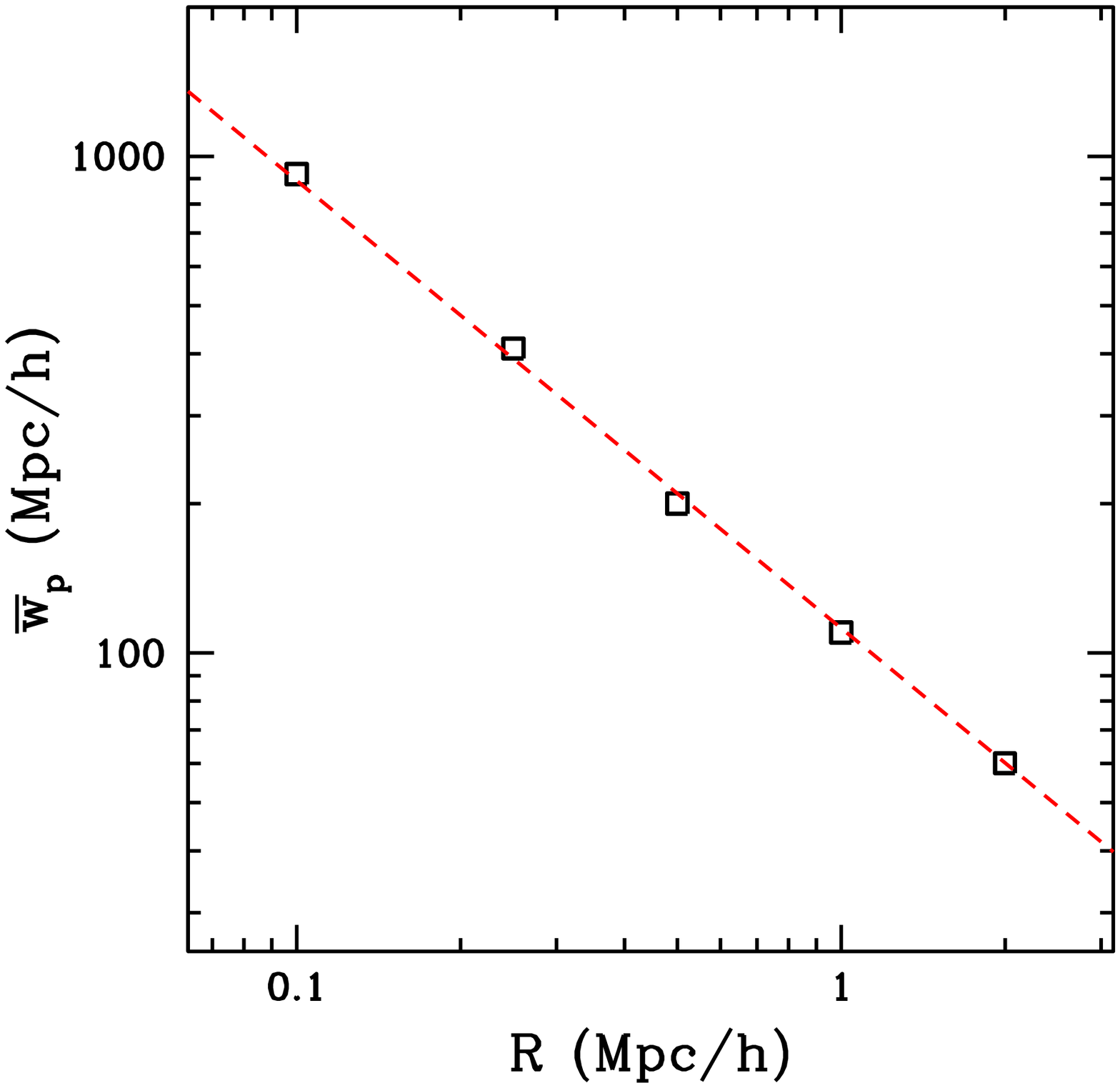}
\end{center}
\caption{The integrated, projected cross-correlation function,
$\bar{w}_{qp}(<R)$, of photometric galaxies and quasars in SDSS from
\protect\citet{Str07} assuming
$f(\chi_q)\approx 10^{-3}\,h\,{\rm Mpc}^{-1}$ (see text). The quoted 
errors are the size of the plotted symbols. The line is derived from
a power law correlation function with slope $1.9$ and $r_{0} = 5\,h^{-1}{\rm Mpc}$.
}
\label{fig:strand}
\end{figure}

A detailed modeling of all these effects goes well beyond the scope of this Appendix.
However, as an illustration, we take the results of \cite{Str07}, subtract 1,
and divide by $\langle f(\chi) \rangle \sim 10^{-3} h {\rm Mpc}^{-1}$; the results are
in Fig.~\ref{fig:strand}. 
The clustering is consistent with a three dimensional power law of slope $1.9$
and $r_{0} \sim 5\,h^{-1}{\rm Mpc}$. We remind the reader that the value of $r_{0}$ is completely
degenerate with our assumption for $\langle f(\chi) \rangle$.

\end{document}